\documentclass[11pt]{article}
\pdfoutput=1
\usepackage{float,jheppub,slashed}
\usepackage{braket,commath}
\usepackage{subcaption,siunitx}
\usepackage{dcolumn}
\usepackage{multirow}
\usepackage{stackengine}
\newcommand\xrowht[2][0]{\addstackgap[.5\dimexpr#2\relax]{\vphantom{#1}}}

\newcommand{\cO}{\mathcal{O}}
\newcommand{\cM}{\mathcal{M}}
\newcommand{\cA}{\mathcal{A}}
\newcommand{\cB}{\mathcal{B}}
\newcommand{\cC}{\mathcal{C}}
\newcommand{\pX}{\mathcal{X}}
\newcommand{\pY}{\mathcal{Y}}
\newcommand{\dd}{{\mathrm d}}
\newcommand{\tr}{\mathrm{tr}}

\title{Quantum Entanglement and Bell Inequality Violation in Semi-Leptonic Top Decays}
\author{Tao Han,}
\author{Matthew Low,}
\author{and Tong Arthur Wu}
\affiliation{PITT PACC, Department of Physics and Astronomy,\\ University of Pittsburgh, 3941 O’Hara St., Pittsburgh, PA 15260, USA}

\emailAdd{than@pitt.edu}
\emailAdd{mal431@pitt.edu}
\emailAdd{tow39@pitt.edu}

\preprint{
\begin{flushright}
PITT-PACC-2316
\end{flushright}}

\abstract{
Quantum entanglement is a fundamental property of quantum mechanics.  Recently, studies have explored entanglement in the $t\bar{t}$ system at the Large Hadron Collider (LHC) when both the top quark and anti-top quark decay leptonically.  Entanglement is detected via correlations between the polarizations of the top and anti-top and these polarizations are measured through the angles of the decay products of the top and anti-top.  In this work, we propose searching for evidence of quantum entanglement in the semi-leptonic decay channel where the final state includes one lepton, one neutrino, two $b$-flavor tagged jets, and two light jets from the $W$ decay.  We find that this channel is both easier to reconstruct and has a larger effective quantity of data than the fully leptonic channel.  As a result, the semi-leptonic channel is $60\%$ more sensitive to quantum entanglement and a factor of 3 more sensitive to Bell inequality violation, compared to the leptonic channel.  In $139~{\rm fb}^{-1}$ ($3~{\rm ab}^{-1}$) of data at the LHC (HL-LHC), it should be feasible to measure entanglement at a precision of $\lesssim 3\%\ (0.7\%)$.  Detecting Bell inequality violation, on the other hand, is more challenging.  With $300~{\rm fb}^{-1}$ ($3~{\rm ab}^{-1}$) of integrated luminosity at the LHC Run-3 (HL-LHC), we expect a sensitivity of $1.3\sigma$ ($4.1 \sigma$).  In our study, we utilize a realistic parametric fitting procedure to optimally recover the true angular distributions from detector effects.  Compared to unfolding this procedure yields more stable results.}

\begin{document}

\maketitle
\flushbottom

%%%%%%%%%%%%%%%%%%%%%%%%%%%%%%%%%%%%%%%%%%%%%%%%%%%%%%%%%%%%%%%%%%%%%%
%%%%%%%%%%%%%%%%%%%%%%%%%%%%%%%%%%%%%%%%%%%%%%%%%%%%%%%%%%%%%%%%%%%%%%
\section{Introduction}
\label{sec:introduction}

Quantum mechanics is at the foundation of modern physics.  One of the novel features of a quantum mechanical system is that it can exhibit entanglement between sub-systems.  Entanglement is a correlation between sub-systems where properly describing one sub-system requires knowledge of the other sub-system, even when the sub-systems are space-like separated.

Another landmark in the understanding of quantum mechanics was the discovery of Bell inequalities~\cite{Bell:1964kc}.  These are inequalities that are satisfied in any classical theory or, more generally, in any local theory that can include hidden variables.  Violations of Bell inequalities, so-called Bell non-localities, indicate that a local classical theory cannot be used to describe these phenomena.  Observations of violations of Bell inequalities are among the strongest experimental evidence for quantum mechanics.

High energy particle colliders fundamentally rely on quantum field theory for their quantitative description and aspects of quantum mechanics are observable throughout the theoretical and experimental landscape.  For instance, interference effects in production cross sections and detection methods for particles rely on quantum mechanics, while precision physics depends on higher-order quantum corrections from all relevant energy scales.  In recent work, the final state in a collider is cast as a system of two qubits which allows us to perform a number of experiments using this system.  Treating the outgoing particles at a collider as a quantum state is a novel experiment that measures and tests quantum mechanics in an unprecedented high-energy regime, many orders of magnitude in energy above conventional quantum experiments.

Adapting to the collider environment presents interesting challenges as there is much less control over the experimental set-up.  On the other hand, at a collider there is an enormous amount of data collected, a wide range of kinematics and energies are explored, and effects that are enhanced at higher energies, like higher dimensional operators, may be visible~\cite{Aoude:2022imd,Severi:2022qjy}.

Recently, there has been a growing body of work on the $t\bar{t}$ system as a quantum state.  First, it was shown that in the fully leptonic channel, where both the top and the anti-top decay leptonically, entanglement could be measured at the Large Hadron Collider (LHC) when only events near threshold are used~\cite{Afik:2020onf}.  Initially, it was predicted that Bell inequality violation, using the same spin correlation observables, could be probed at the high luminosity LHC (HL-LHC)~\cite{Fabbrichesi:2021npl}.  However, subsequent studies found the expected significance to be less than $2\sigma$ when unbiased observables are used~\cite{Severi:2021cnj}.  Ref.~\cite{Afik:2022kwm} noted that one could use expectation values of spin correlations rather than spin correlations themselves to identify entanglement and Bell inequality violation.  Additional significance may be gained by directly measuring an observable sensitive Bell inequality violation, rather than first reconstructing the quantum state and then computing observables from it~\cite{Aguilar-Saavedra:2022uye}.  Beyond Bell inequality violation, other quantum properties can be studied in the $t\bar{t}$ system like quantum steering and quantum discord~\cite{Afik:2022dgh}.  

The issue of spin correlations at colliders is a well-studied topic.  The $t\bar{t}$ system, in particular, has been studied since before the LHC era~\cite{Barger:1988jj,Mahlon:1995zn,Stelzer:1995gc,Parke:1996pr,Mahlon:1997uc,Mahlon:2010gw}.  What is new in the current iteration of work is carefully casting the $t\bar{t}$ system into a quantum state rather than just correlations between two spins.  This allows us to make quantitative statements about the quantum aspects of the $t\bar{t}$ system. 

In this work, we continue the study of the $t\bar{t}$ final state, but instead of studying the leptonic channel, we consider the semi-leptonic channel where either the top or anti-top decays leptonically and the other decays to a light quark and anti-quark.  One of the nice features of the top (or anti-top)  decaying leptonically is that the lepton (or anti-lepton) carries the maximal amount of information about the top polarization.  In hadronic decays,  some of that information is typically lost.  On the other hand, the branching fraction to the semi-leptonic channel is much higher, roughly about a factor of six, so the effective amount of data collected is larger. Combining the more favorable kinematical reconstruction, we find that the semi-leptonic channel is expected to be more sensitive.  While finalizing this work, Ref.~\cite{Dong:2023xiw} presented a study on the semi-leptonic decay of $t\bar{t}$ using unfolding and machine learning for reconstruction. Our work is complementary as we show that choosing an appropriate signal region is impactful and we focus on providing intuition through each stage as well as the theoretical underpinnings.  Instead of unfolding, we utilize parametric fitting. 

In addition to the $t\bar{t}$ system, there have been studies on quantum properties of other systems at colliders.  These include entanglement between two vectors~\cite{Barr:2022wyq,Fabbrichesi:2023cev,Aoude:2023hxv,Bi:2023uop} including production from $h \to VV$~\cite{Barr:2021zcp,Aguilar-Saavedra:2022mpg,Aguilar-Saavedra:2022wam,Bernal:2023ruk,Ashby-Pickering:2022umy,Fabbrichesi:2023jep} and vector boson fusion~\cite{Morales:2023gow}, between  $W$ and $t$~\cite{Aguilar-Saavedra:2023hss}, between $\tau^+$ and $\tau^-$\cite{Altakach:2022ywa,Ma:2023yvd}, between $B$-mesons~\cite{Takubo:2021sdk}, and others~\cite{Fabbrichesi:2022ovb}.  Implications for higher dimensional operators have been explored~\cite{Aoude:2022imd,Severi:2022qjy}, as have other quantum properties like discord and steering~\cite{Afik:2022dgh}.\footnote{This was also studied in the 1990's for $e^+ e^- \to \tau^+ \tau^-$~\cite{Privitera:1991nz,Dreiner:1992gt,Abel:1992kz}. In Sec.~\ref{sec:bell} we reconcile these past works with our work.}

The rest of the paper is organized as follows. In Sec.~\ref{sec:qm}, we review the basics of quantum mechanics with an emphasis on entanglement and Bell inequality violation.  We discuss the general features and the quantum mechanical aspects of the $t\bar{t}$ system for production and decay at hadron colliders in Sec.~\ref{sec:ttbar}.  The results of our analysis on the sensitivity to test entanglement and Bell inequality violation in the $t\bar{t}$ system at the LHC are presented in Sec.~\ref{sec:lhc}.  In Sec.~\ref{sec:conclusion}, we summarize our study, compare with the existing literature, and draw our conclusions.  Some technical aspects of our treatment are included in a few appendices, including a description of different unfolding methods and parametric fitting in Appendix~\ref{app:unfolding}, numerical comparisons with past works in Appendix~\ref{app:comparison}, a presentation of the spin analyzing power for hadronic top decays in Appendix~\ref{app:spin}, a discussion on the fictitious states adopted for detecting entanglement and Bell inequality violation at collider in Appendix~\ref{app:qstate}. Finally,  the potential of charm tagging is  covered in Appendix~\ref{app:ctagging}.

%%%%%%%%%%%%%%%%%%%%%%%%%%%%%%%%%%%%%%%%%%%%%%%%%%%%%%%%%%%%%%%%%%%%%%
%%%%%%%%%%%%%%%%%%%%%%%%%%%%%%%%%%%%%%%%%%%%%%%%%%%%%%%%%%%%%%%%%%%%%%
\section{Quantum Mechanics}
\label{sec:qm}

In this section we first review a few relevant aspects of quantum mechanics, then we discuss entanglement and Bell inequalities.

%%%%%%%%%%%%%%%%%%%%%%%%%%%%%%%%%%%%%%%%%%%%%%%%%%%%%%%%%%%%%%%%%%%%%%
\subsection{Review}
\label{sec:review}

Consider a bipartite system of two qubits.  There is one qubit $|\psi_\cA\rangle$ from sub-system $\cA$ and one qubit $|\psi_\cB\rangle$ from sub-system $\cB$.  These states are vectors in the Hilbert spaces $\mathcal{H}_\cA$ and $\mathcal{H}_\cB$, respectively.  The bipartite state is a vector in the Hilbert space $\mathcal{H}_\cA \otimes \mathcal{H}_\cB$.

A density matrix $\rho$ is a non-negative operator on Hilbert space.  For a state vector $|\psi\rangle$, the associated density matrix is the projection operator $\rho = |\psi\rangle\langle\psi|$.  We will often call $\rho$ itself a quantum state associated with the state vector $|\psi\rangle$.  After choosing a basis, $\rho$ for a bipartite qubit state can be written as a $4 \times 4$ positive semi-definite matrix.

The density matrix formalism is required because it allows us to describe mixed states where state vectors restrict us to pure states.  A mixed state is generically written as
\begin{equation} \label{eq:mixed}
\rho_{\rm mixed} = \sum_{a=1}^N p_a \rho_a,
\quad\quad\quad
\sum_{a=1}^N p_a = 1,
\end{equation}
where $p_a$ is the fraction of the ensemble for the sub-state $a$.  The case of $N=1$ is a pure state, otherwise, it is a mixed state.  In our application to the $t\bar{t}$ system, we will be dealing with a mixed state.

For a single qubit, the density matrix can be described by the Pauli decomposition
\begin{equation} \label{eq:PauliDecomposition2}
\rho = \frac{1}{2} \Big( \mathbb{I}_2 + \sum_i B_i \sigma_i \Big),
\end{equation}
where $\sigma_i\ (i=1,2,3)$ are the Pauli matrices and $B_i$ are the corresponding vector components describing the net polarization of the qubit.  A bipartite qubit system follows the Pauli decomposition in a similar way
\begin{equation} \label{eq:PauliDecomposition4}
\rho = \frac{1}{4} \Big( \mathbb{I}_4 
+ \sum_i \big(B^\cA_i \; (\sigma_i \otimes \mathbb{I}_2)
+ B^\cB_i \; (\mathbb{I}_2 \otimes \sigma_i)\big)
+ \sum_{i,j} C_{ij} \; (\sigma_i \otimes \sigma_j) \Big).
\end{equation}
For a general state, there are $3+3+9=15$ degrees of freedom from the vectors $B^\cA_i$ and $B^\cB_i$, and the matrix $C_{ij}$. The $B^\cA_i$ vector is the net polarization of spin $\cA$, the $B^\cB_i$ vector is the net polarization of spin $\cB$, and $C_{ij}$ is the spin correlation matrix between sub-systems $\cA$ and $\cB$.  In many cases of interest, some of these parameters are zero by symmetry.

Determining all the parameters $\{ B^\cA_i, B^\cB_j, C_{ij} \}$ implies that $\rho$ can be reconstructed, which is known as quantum tomography. Once the quantum state $\rho$ has been measured, the expectation value of any  observable $\cO$ can be computed as
\begin{equation} \label{eq:expectationValue}
\langle \cO \rangle = \tr( \cO \rho ).
\end{equation}
For instance, the net polarization of qubit $\cA$ corresponds to the operator $\cO = \sigma_i \otimes \mathbb{I}_2$. By Eqs.~\eqref{eq:PauliDecomposition4} and~\eqref{eq:expectationValue}, this is $\langle \sigma_i \otimes \mathbb{I}_2 \rangle = B^\cA_i$.

%%%%%%%%%%%%%%%%%%%%%%%%%%%%%%%%%%%%%%%%%%%%%%%%%%%%%%%%%%%%%%%%%%%%%%
\subsection{Entanglement}
\label{sec:entanglement}

Consider a state $\rho$ for a bipartite system with sub-systems $\cA$ and $\cB$.  This state is separable if it can be written as a factorized product 
\begin{equation}
\rho = \sum_{a=1}^N p_a \; \rho^\cA_a \otimes \rho^\cB_a.
\end{equation}
If it cannot be written in this separable factorized form, it is entangled.  This means that sub-system $\cA$ cannot be fully described without knowledge of sub-system $\cB$.  For a pure state, $N=1$ and $p_1=1$.

Given a state $\rho$ there are different ways to determine if $\rho$ describes an entangled or a separable state.  We choose to use the Peres-Horodecki criterion, also called the positive partial transpose (PPT) criterion~\cite{Peres:1996dw,Horodecki:1997vt}.  The PPT criterion performs the transpose on sub-system $\cB$ and leaves sub-system $\cA$ unmodified leading to a matrix $\rho^{T_\cB}$ where
\begin{equation}
\rho^{T_\cB} = (\mathbb{I}_2 \otimes T_\cB) \rho .
\end{equation}
The matrix $\rho^{T_\cB}$ may or may not be a state.  For a separable, unentangled state $\rho_{\rm sep}$, the associated $\rho_{\rm sep}^{T_\cB}$ can be written as $\sum_a p_a \; \rho^\cA_a \otimes (\rho^\cB_a)^T$, which corresponds to a valid state.  For an entangled state $\rho_{\rm ent}$, however, the associated $\rho_{\rm ent}^{T_\cB}$ is no longer a state.

In general, a matrix is a valid state if all of its eigenvalues are $\geq 0$, or equivalently stated, the matrix is positive semi-definite.  The PPT criterion leads to a list of inequalities, the violation of any of these inequalities is a sufficient, but not necessary condition for entanglement.  Thus, using the PPT criterion to show entanglement requires just finding a single inequality that isn't satisfied, while showing separability requires checking a set of inequalities.

Concretely, expanding a quantum state $\rho$ according to Eq.~\eqref{eq:PauliDecomposition4} allows us to write the conditions in terms of elements of the spin correlation matrix $C_{ij}$.  The quantum state $\rho$ is
\begin{equation}{\tiny
\rho=\frac{1}{4}
\left(\begin{array}{cccc}
1+B_3^\cA+B_3^\cB+C_{33} & B_1^\cB+C_{31}-i(B_2^\cB+C_{32})& B_1^\cA+C_{13}-i(B_2^\cA+C_{23})&
C_{11}-C_{22}-i(C_{12}+C_{21})\cr
B_1^\cB+C_{31}+i(B_2^\cB+C_{32})&
1+B_3^\cA-B_3^\cB-C_{33}&
C_{11}+C_{22}+i(C_{12}-C_{21})&
B_1^\cA-C_{13}-i(B_2^\cA-C_{23})\cr
B_1^\cA+C_{13}+i(B_2^\cA+C_{23})&
C_{11}+C_{22}-i(C_{12}-C_{21})&
1-B_3^\cA+B_3^\cB-C_{33}&
B_1^\cB-C_{31}-i(B_2^\cB-C_{32})\cr
C_{11}-C_{22}+i(C_{12}+C_{21})&
B_1^\cA-C_{13}+i(B_2^\cA-C_{23})&
B_1^\cB-C_{31}+i(B_2^\cB-C_{32})&
1-B_3^\cA-B_3^\cB+C_{33}
\end{array}\right),
}\end{equation}
and the matrix $\rho^{T_\cB}$ is
\begin{equation}{\tiny
\rho^{T_\cB}=\frac{1}{4}
\left(\begin{array}{cccc}
1+B_3^\cA+B_3^\cB+C_{33} & B_1^\cB+C_{31}+i(B_2^\cB+C_{32})& B_1^\cA+C_{13}-i(B_2^\cA+C_{23})&
C_{11}+C_{22}+i(C_{12}-C_{21})\cr
B_1^\cB+C_{31}-i(B_2^\cB+C_{32})&
1+B_3^\cA-B_3^\cB-C_{33}&
C_{11}-C_{22}-i(C_{12}+C_{21})&
B_1^\cA-C_{13}-i(B_2^\cA-C_{23})\cr
B_1^\cA+C_{13}+i(B_2^\cA+C_{23})&
C_{11}-C_{22}+i(C_{12}+C_{21})&
1-B_3^\cA+B_3^\cB-C_{33}&
B_1^\cB-C_{31}+i(B_2^\cB-C_{32})\cr
C_{11}+C_{22}-i(C_{12}-C_{21})&
B_1^\cA-C_{13}+i(B_2^\cA-C_{23})&
B_1^\cB-C_{31}-i(B_2^\cB-C_{32})&
1-B_3^\cA-B_3^\cB+C_{33}
\end{array}\right).
}\end{equation}
One example of a sufficient condition for entanglement can be derived from deleting the 2$^{nd}$ and 3$^{rd}$ rows and columns of this matrix \cite{Afik:2020onf}, leading to 
\begin{equation} \label{eq:inequalityForEntanglement}
|C_{11} + C_{22}| > 1 + C_{33}.
\end{equation}
Whether $(C_{11}+C_{22})$ is positive or negative leads to two separate cases of $C_{11} + C_{22} > 1 + C_{33}$ and $-C_{11} - C_{22} > 1 + C_{33}$.  Rearranging these inequalities we write $O_E^\pm = \pm C_{11} \pm C_{22} - C_{33} - 1$ where $O_E^\pm > 0$ indicates entanglement.  It will be shown in Sec.~\ref{sec:collider} that the quantity $O_E^\pm$ corresponds to an observable $\cO_E^\pm$ such that testing entanglement at a collider becomes
\begin{equation} \label{eq:sampleOpEnt}
\cO_E^\pm = \pm C_{11} \pm C_{22} - C_{33} - 1 , 
\quad\quad {\rm and} \quad\quad
\langle \cO_E^\pm \rangle > 0\quad 
{\rm for\  entanglement.} 
\end{equation}
In pre-defined regions, the observable $\cO_E^\pm$ corresponds to whether the quantum state $\rho$ is entangled or separable.

It is also customary to introduce a quantity called the ``concurrence'' $\cC$~\cite{Wootters:1997id}, which is defined for bipartite qubit systems as 
\begin{equation} \label{eq:def-concurrence}
\cC(\rho) = \max(0,\lambda_1 - \lambda_2 - \lambda_3 - \lambda_4),
\end{equation}
where $\lambda_i$ ($i=1,2,3,4$) are the eigenvalues, sorted by decreasing magnitude, of the matrix 
\begin{equation}
R_\rho = \sqrt{\sqrt{\rho} \tilde{\rho} \sqrt{\rho}}, 
\quad\quad\quad 
\tilde{\rho} = (\sigma_2 \otimes \sigma_2) \rho^* (\sigma_2 \otimes \sigma_2).
\end{equation}
For a separable state $\rho_{\rm sep}$, the concurrence is $\cC(\rho_{\rm sep})=0$, while for an entangled state $\rho_{\rm ent}$ the concurrence is $0 < \cC(\rho_{\rm ent}) \leq 1$.\footnote{For intuition, consider the simplified case when $\rho$ is a pure state.  The concurrence $\cC$ can be written as $\tr(\rho_A^2) = 1 - \cC^2/2$ where $\rho_A$ is the reduced density matrix obtained by taking the partial trace with respect to sub-system $\cB$ of $\rho$.  The concurrence then measures how far the reduced density matrix is from a pure state.  Generalizing this to mixed states leads to Eq.~\eqref{eq:def-concurrence}.} 

Therefore one method of identifying entanglement is to first fully determine $\rho$, and then compute $\cC(\rho)$ to be zero or not.  It can be shown that in the $t\bar{t}$ system Eq.~\eqref{eq:sampleOpEnt} is equal to the concurrence.

%%%%%%%%%%%%%%%%%%%%%%%%%%%%%%%%%%%%%%%%%%%%%%%%%%%%%%%%%%%%%%%%%%%%%%
\subsection{Bell Inequality Violation}
\label{sec:bell}

By construction, a Bell inequality holds for any system that can be described by a local hidden variable theory~\cite{Bell:1964kc}.  Bell inequality violation indicates that a given theory must be either classically non-local, or quantum-mechanically entangled.  This historically was very strong evidence for quantum mechanics.  A separable state always satisfies Bell's inequality, while an entangled quantum state may or may not violate a Bell inequality. Therefore, Bell inequality violation is a stricter test of ``quantumness'' than entanglement.

For a bipartite system of two qubits, the only Bell inequality is the CHSH inequality (Clauser-Horne-Shimony-Holt)~\cite{Clauser:1969ny}, which reads
\begin{equation} \label{eq:CHSH}
\langle A_1 B_1 \rangle 
- \langle A_1 B_2 \rangle
+ \langle A_2 B_1 \rangle
+ \langle A_2 B_2 \rangle \leq 2.
\end{equation}
The first term is a simultaneous measurement $A_1$ on sub-system $\cA$ and $B_1$ on sub-system $\cB$.  The other terms are measured in a likewise manner.  A quantum state of a bipartite system that violates Eq.~\eqref{eq:CHSH} exhibits Bell inequality violation (or is Bell non-local).

For the case where the two qubits are spins, $A_1$ and $A_2$ can indicate the quantization axes along which the spin of qubit $\cA$ is measured while $B_1$ and $B_2$ can indicate the axes along which the spin of qubit $\cB$ is measured.  For instance the choice of
\begin{equation} 
\label{eq:CHSH2}
A_1 = \sigma_3, \quad\quad
A_2 = \sigma_1, \quad\quad
B_1 = -\frac{1}{\sqrt{2}}(\sigma_1 + \sigma_3), \quad\quad
B_2 = \frac{1}{\sqrt{2}}(\sigma_1-\sigma_3), \quad\quad
\end{equation}
when applied to the Bell state $\psi_{\rm Bell} = (|01\rangle - |10\rangle)/\sqrt{2}$ violates the CHSH inequality.

Given a quantum state, it is crucial to choose the optimal axes in order to determine if a quantum state violates the CHSH inequality, via Eq.~\eqref{eq:CHSH}.  It has been shown that while using the optimal axes, the left-hand side of Eq.~\eqref{eq:CHSH} becomes $2\sqrt{\lambda_1 + \lambda_2}$, where $\lambda_1$ and $\lambda_2$ are the two largest eigenvalues of $C^T C$~\cite{Horodecki1995ViolatingBI}.  In a collider environment, however, this method can lead to a biased estimation~\cite{Fabbrichesi:2021npl,Severi:2021cnj}.

For simplicity we will choose the fixed axes~\cite{Aguilar-Saavedra:2022uye}
\begin{equation} \label{eq:CHSHaxes}
A_1 = \sigma_3, \quad\quad
A_2 = \sigma_1, \quad\quad
B_1 = \pm \frac{1}{\sqrt{2}}(\sigma_3 + \sigma_1), \quad\quad
B_2 = \pm \frac{1}{\sqrt{2}}(- \sigma_3 + \sigma_1). \quad\quad
\end{equation}
For this choice the CHSH inequality becomes
\begin{equation} \label{eq:CHSHineq}
|C_{33} \pm C_{11}| < \sqrt{2}.
\end{equation}
In a similar way to entanglement, we can cast this into an observable as
\begin{equation}
\cO_B^\pm = \pm (C_{33} + C_{11}) - \sqrt{2}, 
\quad\quad {\rm and} \quad\quad
\langle \cO_B^\pm \rangle > 0\quad 
\text{for\  Bell inequality violation.} 
\end{equation}
Whether the $+$ or $-$ is used depends on the predicted value of $C_{33} + C_{11}$.

\paragraph{}
Finally, we make a comment about the generality of the Bell inequality violation test that can be performed at a collider.  In the 1990's, it was suggested that Bell inequality violation could be observed at $e^+ e^-$ colliders in the $\tau^+ \tau^-$ final state~\cite{Privitera:1991nz,Dreiner:1992gt,Abel:1992kz}.  The conclusion of Ref.~\cite{Abel:1992kz} was that Bell inequality violation was not observable at a collider because quantities measured at colliders are commuting while non-commuting quantities are required to violate a Bell inequality.  In this work, we do not perform a fully general test of Bell's inequality.  Instead, we first identify a quantum state, and then ask whether it is a quantum state that does or does not violate Bell's inequality.  The non-commutation arises from our assumption that we are working with a quantum state and thus gain access to spins.

%%%%%%%%%%%%%%%%%%%%%%%%%%%%%%%%%%%%%%%%%%%%%%%%%%%%%%%%%%%%%%%%%%%%%%
%%%%%%%%%%%%%%%%%%%%%%%%%%%%%%%%%%%%%%%%%%%%%%%%%%%%%%%%%%%%%%%%%%%%%%
\section{The Top-Antitop System at Hadron Colliders}
\label{sec:ttbar}

In this section, we cover the details that are necessary to identify the $t\bar{t}$ final state at the LHC as a quantum state.

%%%%%%%%%%%%%%%%%%%%%%%%%%%%%%%%%%%%%%%%
\subsection{Two-Body Production at Hadron Colliders} 
\label{sec:collider}

Consider the two-to-two scattering process $\pX \pY \to \cA \cB$.  The rate for this process is given by the cross section $\sigma(\pX\pY \to \cA\cB)$ and is calculated by taking the matrix element $\mathcal{M}(\pX \pY \to \cA \cB)$, squaring it, and integrating it over phase space $\dd\Pi$.  The initial state spins (and other quantum numbers) are averaged, and when the final state spins are not measured they are summed.  Schematically
\begin{equation}
\sigma(\pX\pY \to \cA\cB) = \int \dd\Pi \ 
\overline{\sum_{\rm initial}} \;
\sum_{ab,\bar{a}\bar{b}} 
\cM(\pX\pY \to \cA\cB)_{a\bar{a}} 
\cM^*(\pX\pY \to \cA\cB)_{b\bar{b}},
\end{equation}
where $ab$ is the spin index of particle $\cA$, $\bar{a}\bar{b}$ is the spin index of particle $\cB$, and $\overline{\sum}$ indicates averaging.

The {\it production spin density matrix} is
\begin{equation} \label{eq:productionSpinDensityMatrix}
R_{ab,\bar{a}\bar{b}} =
\overline{\sum_{\rm initial}} 
\cM(\pX\pY \to \cA\cB)_{a\bar{a}} 
\cM^*(\pX\pY \to \cA\cB)_{b\bar{b}},
\end{equation}
such that
\begin{equation}
\sigma(\pX\pY \to \cA\cB) = \int \dd\Pi 
\sum_{ab,\bar{a}\bar{b}} 
R_{ab,\bar{a}\bar{b}}.
\end{equation}
Taking the trace of $R_{ab,\bar{a}\bar{b}}$ and performing the phase space integral gives the cross section, while the full matrix provides differential spin information.
When particles $\cA$ and $\cB$ are both spin-$1/2$, the matrix $R_{ab,\bar{a}\bar{b}}$ is a $4 \times 4$ matrix and can be decomposed into the Pauli basis according to Eq.~\eqref{eq:PauliDecomposition4}.\footnote{The normalization for a production spin density matrix $R$ is $\tr(R) = d\sigma/d\Pi$ while the normalization for a quantum state $\rho$ is $\tr(\rho) = 1$.}

If particle $\cA$ decays, the {\it decay spin density matrix} carries the differential spin information of particle $\cA$.  Consider the three-body decay of $\cA \to a_1 a_2 a_3$
\begin{equation}
    \Gamma^\cA_{ab} = \cM(\cA \to a_1 a_2 a_3)_{a} 
    \cM^*(\cA \to a_1 a_2 a_3)_{b},
\end{equation}
where again $ab$ is the spin index of particle $\cA$.

In the narrow width approximation the production and decay can be described together
\begin{equation}
\sigma(\pX\pY \to \cA\cB \to (a_1 a_2 a_3)(b_1 b_2 b_3)) = 
\int \dd\Pi \; 
\sum_{ab,\bar{a}\bar{b}} (\Gamma^\cA_{ab} \; R_{ab,\bar{a}\bar{b}} \; \Gamma^\cB_{\bar{a}\bar{b}}).
\end{equation}
The final state phase space can be partially integrated over to find
\begin{equation}\begin{aligned}
\int \dd\Pi \; \sum_{ab,\bar{a}\bar{b}}
(\Gamma^\cA_{ab} \; R_{ab,\bar{a}\bar{b}} \; \Gamma^\cB_{\bar{a}\bar{b}})
&=
\int \dd\Omega^\cA \dd\Pi^\cA
\dd\Omega^\cB \dd\Pi^\cB \; \sum_{ab,\bar{a}\bar{b}} 
(\Gamma^\cA_{ab} \; R_{ab,\bar{a}\bar{b}} \; \Gamma^\cB_{\bar{a}\bar{b}}), \\
&=
\int \dd\Omega^\cA \dd\Omega^\cB \; \sum_{ab,\bar{a}\bar{b}} 
(\tilde{\Gamma}^\cA_{ab} \;
R_{ab,\bar{a}\bar{b}} \;
\tilde{\Gamma}^\cB_{\bar{a}\bar{b}}).
\end{aligned}\label{eq:totalintegral}\end{equation}
The total phase space $\dd\Pi$ is divided into the angular phase space of one of the decay products of particle $\cA$:  $\dd\Omega^\cA$, the angular phase space of one of the decay products of particle $\cB$: $\dd\Omega^\cB$, the remaining phase space of the decay products of particle $\cA$: $\dd\Pi^\cA$, and remaining phase space of the decay products of particle $\cB$: $\dd\Pi^\cB$.  The angular space is two-dimensional $(\theta,\phi)$ but we write it as a three-vector $\Omega_i = (\cos\phi\sin\theta,\sin\phi\sin\theta,\cos\theta)$ to represent the direction of the decay product of interest.  Here, $\theta$ is the polar angle and $\phi$ is the azimuthal angle with respect to a reference direction. 

While $\Gamma^\cA_{ab}$ is the decay spin density matrix for particle $\cA$, $\tilde{\Gamma}^\cA_{ab}$ is the partially integrated decay width that leaves the angular space of one of the decay products unintegrated.  It can be decomposed as in Eq.~\eqref{eq:PauliDecomposition2} to $\tilde{\Gamma}^\cA_{ab} \propto \delta_{ab} + \sum_i B^\cA_i \sigma_{i,ab}$ where $B^\cA_i$ is the net polarization of particle $\cA$.  Performing the calculation of $\tilde{\Gamma}^\cA_{ab}$ in the rest frame of particle $\cA$ leads to
\begin{equation} \label{eq:decayspindensity}
\tilde{\Gamma}^A_{ab}(\Omega_i)
= \frac{1}{2} \Gamma^A \Big(
\delta_{ab}
+ \sum_i B^\cA (\kappa \Omega_i) \sigma_{i,ab} \Big),    
\end{equation}
where $\Gamma^\cA$ is proportional to the decay width of $\cA \to a_1 a_2 a_3$, $B^\cA$ is the magnitude of the polarization of particle $\cA$, and $\kappa$ is called the {\it spin analyzing power} and is associated with the decay particle that has been left unintegrated. The value of $\kappa$ is between $-1$ and $1$ and describes how correlated a decay product is with the spin of the mother particle.

Writing the decay spin density matrix according to Eq.~\eqref{eq:decayspindensity}, decomposing the production spin density matrix according to Eq.~\eqref{eq:PauliDecomposition4}, and summing over $ab, \bar{a}\bar{b}$ in Eq.~\eqref{eq:totalintegral}, the differential cross section can be written as 
\begin{equation} \label{eq:observablesToAngles}
\frac{1}{\sigma} \frac{\dd^4\sigma}{\dd^2\Omega^\cA \dd^2\Omega^\cB}
= \frac{1}{(4\pi)^2} 
\bigg(1+\sum_{i}\big(\kappa^\cA \, B^\cA_i \Omega^A_i +
\kappa^\cB \, B^\cB_i \Omega_i^\cB\big) + \sum_{i,j} 
\kappa^\cA \kappa^\cB \, \Omega_i^\cA C_{ij} \Omega_j^\cB\bigg),
\end{equation}
where the angle $\Omega_i^\cA$ ($\Omega_j^\cB$) is evaluated in the rest frame of particle $\cA$ ($\cB$) relative to the $i^{\rm th}$ ($j^{\rm th}$) axis of a chosen basis. 

To extract individual parameters, one can select which angular integrals to perform.  For example, to extract a component of the spin correlation matrix $C_{ij}$, one integrates over $\phi^\cA$ and $\phi^\cB$ to obtain
\begin{equation} \label{eq:cosA_cosB}
\frac{1}{\sigma} \frac{\dd \sigma}{\dd \small\cos{\theta^\cA_i}\, \dd\cos{\theta^\cB_j} } = -\frac{1}{4} \left(1+ \kappa^\cA B_i^\cA \cos{\theta^\cA_i} + \kappa^\cB B_j^\cB \cos{\theta^\cB_j}+ \kappa^\cA \kappa^\cB C_{ij} \cos{\theta^\cA_i}\cos{\theta^\cB_j} \right),
\end{equation}
where $\theta^\cA_i$ ($\theta^\cB_j$) is the angle between the momentum of the decay product of particle $\cA$ ($\cB$) and the $i^{\rm th}$ ($j^{\rm th}$) axis, in the rest frame of particle $\cA$ ($\cB$).

This distribution can be transformed to
\begin{equation} \label{eq:cosAcosB}
\frac{1}{\sigma} \frac{\dd \sigma}{\dd \small(\cos{\theta^\cA_i}\cos{\theta^\cB_j} )} = -\frac{1}{2}\left(1+\kappa^\cA \kappa^\cB C_{ij} \cos{\theta^\cA_i}\cos{\theta^\cB_j} \right)
\log \big|\cos{\theta^\cA_i}\cos{\theta^\cB_j} \big| ,
\end{equation}
Thus measuring angles of decay products measures parameters of the production spin density matrix.

We mention three ways to extract the value of $C_{ij}$ from data using Eq.~\eqref{eq:cosAcosB}.  The first way is to simply perform a fit to the differential cross section.

The second way is to compute the asymmetry of the distribution.  The asymmetry $A$ for a variable $x$ is
\begin{equation} \label{eq:asymmetry}
A_x = \frac{N_x^+ - N_x^-}{N_x^+ + N_x^-},
\end{equation}
where $N_x^+$ ($N_x^-$) is the number of events with $x>0$ ($x<0$):
\begin{equation}
N_x^+ = \int_0^{x_{\rm max}} \frac{1}{\sigma} \frac{\dd \sigma}{\dd x} \dd x, 
\quad\quad\quad
N_x^- = \int_{x_{\rm min}}^0 \frac{1}{\sigma} \frac{\dd \sigma}{\dd x} \dd x.
\end{equation}
When the asymmetry variable is $x=\cos{\theta^\cA_i}\cos{\theta^\cB_j}$ then $x_{\rm max}=1$ and $x_{\rm min}=-1$.  This method works because $C_{ij}$ multiplies the component of the differential cross section that is an odd function with respect to $\cos{\theta^\cA_i}\cos{\theta^\cB_j}$.

Each spin correlation matrix entry $C_{ij}$ is then
\begin{equation} \label{eq:cij_individual}
C_{ij} = \frac{4}{\kappa^\cA \kappa^\cB}
\left( A_{\cos{\theta^\cA_i}\cos{\theta^\cB_j}} \right).
\end{equation}
The third way to extract $C_{ij}$ is to compute the mean of the distribution in Eq.~\eqref{eq:cosAcosB} since $\langle \cos{\theta^A_i}\cos{\theta^B_j} \rangle~\propto~C_{ij}$, where the constant of proportionality depends on the distribution.  The variance of the mean is smaller than the variance of the asymmetry, however, the asymmetry is more robust to systematic uncertainties.  In our study we utilize the asymmetry.

%%%%%%%%%%%%%%%%%%%%%%%%%%%%%%%%%%%%%%%%%%%%%%%%%%%%%%%%%%%%%%%%%%%%%%
\subsection{The \texorpdfstring{$t\bar{t}$}{tt} System as a Quantum State}
\label{sec:ttsystem}

Consider the $t\bar{t}$ final state as a bipartite qubit system with the spin of the top and anti-top identified as each qubit.  Then each event at the LHC is a single measurement of this quantum state.  Each event can also be called a quantum sub-state.  Let the quantum state that describes the $t\bar{t}$ system be $\rho$.

The kinematics of the $t\bar{t}$ system are characterized by the invariant mass,  $m_{t\bar{t}}$, of the top-anti-top pair and by the angle, $\theta$, of the top momentum (in the $t\bar{t}$ center-of-mass frame) relative to the beam.  The quantum state for a single point in this phase space is $\rho(m_{t\bar{t}},\theta)$, while more generally, integrating over a region $\Pi$ leads to the quantum state $\rho_\Pi$~\cite{Afik:2020onf}.

At a hadron collider, the two partonic processes, at leading order, that produce $t\bar{t}$ are $q\bar{q}$ and $gg$.  This means that $\rho$ is necessarily a mixed state where the coefficients, as in Eq.~\eqref{eq:mixed}, are given by the relative parton luminosities \cite{Afik:2020onf}.  Additionally, we can identify the production spin density matrix, Eq.~\eqref{eq:productionSpinDensityMatrix}, as a quantum density matrix for a sub-state (for a given initial partonic state) when normalized correctly and evaluated in a fixed basis~\cite{Afik:2020onf}.

The ideal final state would be the exclusive production of $t\bar{t}$ since additional radiation can disrupt the spin correlations between the $t$ and $\bar{t}$.  In this study, we work at leading order and leave higher order effects to future work.  In the context of spin correlations, higher order effects have been studied and are known to modify spin correlations at the $10-30\%$ level~\cite{Bernreuther:2004jv}.

The main backgrounds for $t\bar{t}$ in the semi-leptonic channel are single top, $W+$ jets, multijet, $t\bar{t} W$, $t\bar{t} Z$, and $t\bar{t} h$.  Altogether the background has a cross section that is $\approx 10\%$ of the size of the signal when two $b$-tags are required~\cite{ATLAS:2020aln,CMS:2023qyl}.  In the boosted region this reduces to $\approx 4\%$~\cite{ATLAS:2022xfj,CMS:2021vhb}.  In this work the impact of backgrounds is neglected and left to future work.

In Bell inequality tests,  loopholes often exist and the $t\bar{t}$ system is no exception.  In some events, the top and anti-top decay while inside of each other's light cones.  This is an example of the locality loophole.  As the invariant mass $m_{t\bar{t}}$ of the $t\bar{t}$ system increases, the fraction of events which are space-like separated when decaying approaches 100\% and is already at 90\% for $m_{t\bar{t}}>800~{\rm GeV}$~\cite{Severi:2021cnj}.  Another loophole is the fair sampling loophole which asserts that if the detection efficiency is low then a violation of Bell's inequality could be faked.  The fair sampling loophole, as well as others, are expected to be difficult to address at colliders.

%%%%%%%%%%%%%%%%%%%%%%%%%%%%%%%%%%%%%%%%%%%%%%%%%%%%%%%%%%%%%%%%%%%%%%
\subsection*{Spin Correlations}

The $t\bar{t}$ system has been studied for many years.  In 1988 it was known that when produced via the strong force ($gg$ and $q\bar{q}$), neither the tops nor anti-tops are polarized at leading order, but that spin correlations exist between the top and anti-top~\cite{Barger:1988jj}.  Furthermore, these spin correlations can be observed by the angular separations between the top and anti-top decay products~\cite{Mahlon:1995zn,Stelzer:1995gc}.  The $gg$ and $q\bar{q}$ initial states give rise to different spin correlation behavior which is also the reason that the LHC and the Tevatron are very complementary probes of this system.  

The heuristic intuition for the spin correlations is that near threshold the spins of the top and anti-top are aligned along the beamline direction.  The possible outgoing spin configurations are controlled by the incoming spins.  For the $q\bar{q}$ initial state the $q$ and $\bar{q}$ have opposite helicity with the spins aligned along the beam axis.  Near threshold the top and anti-top have mostly opposite helicity with spins aligned along the beam axis, leading to a configuration with a spin-triplet contribution.  At high $p_T$, the top and anti-top are still opposite in helicity but their spin axes become aligned with their direction of motion.  In between the threshold region and the high $p_T$ region, the spin axes of the top and anti-top interpolate between these directions~\cite{Parke:1996pr,Mahlon:1997uc}.  The basis for choosing the spin axes is called the off-diagonal basis and has been shown to optimize the spin correlations from $q\bar{q}$ production.

The situation is different for $gg$ production.  Incoming pairs of gluons can have the same helicity or the opposite helicity.  Near threshold same-helicity gluons dominate and the outgoing top and anti-top have the same helicity with the spin axes aligned along the beam axis, leading to a configuration with a spin-singlet contribution,  in contrast to the $q\bar q$ case.  At high $p_T$, the opposite-helicity gluons dominate and the outgoing top and anti-top have opposite helicity with the spin axes becoming aligned along their direction of motion, leading to a spin-triplet configuration, which is the same as the $q\bar q$ case.  The optimal choice of spin axes for optimizing spin correlations is aligned along the direction of the top and anti-top ~\cite{Mahlon:1997uc,Uwer:2004vp,Mahlon:2010gw}.

%%%%%%%%%%%%%%%%%%%%%%%%%%%%%%%%%%%%%%%%%%%%%%%%%%%%%%%%%%%%%%%%%%%%%%
\subsection*{Basis Choice}

\begin{figure} [tb]
  \begin{center}
  \includegraphics[width=0.5\textwidth] {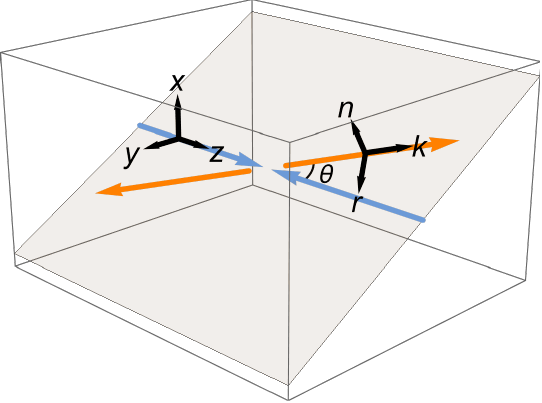}
  \caption{Illustration of the fixed beam basis $\{ \hat{x}, \hat{y}, \hat{z} \}$ and the helicity basis $\{ \hat{r}, \hat{k}, \hat{n} \}$.  The incoming beams are blue and the outgoing top and anti-top are orange. $\theta$ is the polar angle between the $\hat{z}$-axis (the beam direction) and the $\hat{k}$-axis (the top quark momentum direction).}
  \label{fig:bases_diagram}
  \end{center}
\end{figure}

When calculating spin correlations, it is necessary to choose a basis along which to measure the spins.  One common choice is the fixed beam basis which starts from the center-of-mass frame of the $t\bar{t}$ system and uses $\{ \hat{x}, \hat{y}, \hat{z} \}$ where $\hat{z}$ points along the beam and $\hat{x}$ and $\hat{y}$ are fixed in the plane transverse to the beam.
Another common choice is the helicity basis which starts from the center-of-mass frame and defines $\{ \hat{r}, \hat{k}, \hat{n} \}$ where $\hat{k}$ points along the top quark three-momentum.  Then
\begin{subequations}\begin{align}
    \hat{r} &= \frac{1}{\sin\theta}(\hat{z} - \cos\theta \hat{k}), \\
    \hat{n} &= \hat{r} \times \hat{k},
\end{align}\end{subequations}
where $\hat{r}$ is the component of the beam direction that is orthogonal to $\hat{k}$, $\hat{n}$ is the remaining orthogonal direction, and $\cos\theta=\hat{k}\cdot\hat{z}$.  Figure~\ref{fig:bases_diagram} illustrates these two bases.\footnote{We calculate angles for both the top and anti-top using the axes $\{ \hat{r}, \hat{k}, \hat{n} \}$ where $\hat{k}$ is defined by the top quark.  Sometimes in other studies the angles for decay products from the anti-top are defined relative to a second set of axes defined by the anti-top.}

%%%%%%%%%%%%%%%%%%
%%%%%%%%%%%%%%%

At the LHC, at threshold the fixed beam basis has the largest spin correlations while in the high-$p_T$ regime, the helicity basis has the largest correlations~\cite{Bernreuther:2008ju}.  At high-$p_T$ the helicity basis is nearly optimal for entanglement and Bell inequality violation too~\cite{Kun}.

One very important note is that constructing the total $t\bar{t}$ quantum state ({\it i.e.} quantum tomography) requires a fixed basis for all of the events \cite{Afik:2020onf,Afik:2022kwm,Kun}.  In a fixed basis, the axes are the same for each event which means that each event is a single measurement of a parameter of Eq.~\eqref{eq:PauliDecomposition4}.  Using many events then increases the accuracy of the measurement of the $t\bar{t}$ quantum state.  

The helicity basis, by contrast, is not a fixed basis because the axes change event-by-event.  Performing the summation over many events does not measure a parameter of Eq.~\eqref{eq:PauliDecomposition4} but rather its expectation value~\cite{Afik:2022kwm,Kun} since the basis is different event-by-event.  Thus in the helicity basis, the summation over events does not produce a quantum state, but is simply a summation over events.  In Ref.~\cite{Afik:2022kwm} this sum was labelled a ``fictitious state.''  

Showing that a fictitious state is entangled does not show that the associated $t\bar{t}$ quantum state is entangled, but it does show that there exists a sub-state (both of the fictitious state and of the associated quantum state) that is entangled.  
This follows from the fact that both the quantum state and fictitious states are convex sums and the positivity of concurrence.  The same considerations apply to Bell inequality violation.  Appendix~\ref{app:qstate} provides a proof of this statement, as well as more discussion on fictitious states (Ref.~\cite{Kun} provides additional details).  

In our study we use the helicity basis for both concurrence and CHSH violation which means our results indicate the presence of entanglement and of Bell inequality violation, but not the strength.  In the high-$p_T$ region, these are naively not detectable using the fixed beam basis.

%%%%%%%%%%%%%%%%%%%%%%%%%%%%%%%%%%%%%%%%%%%%%%%%%%%%%%%%%%%%%%%%%%%%%%
\subsection{Spin Analyzing Power}
\label{sec:spin}

As seen in Eq.~\eqref{eq:cosAcosB} the measured value of spin correlations is impacted linearly by the spin analyzing power $\kappa^\cA$ from the decay of particle $\cA$ and the spin analyzing power $\kappa^\cB$ from the decay of particle $\cB$.  To maximize the sensitivity and significance, the daughter particle with the largest spin analyzing power should be used.

When the top decays leptonically, the anti-lepton ($\ell^+$) has $\kappa = 1.00$ which is maximally correlated with the spin of the top quark.  In the fully leptonic channel of the $t\bar{t}$ system the lepton and the anti-lepton are used which results in maximal correlation.  In the semi-leptonic channel, that we study here, one side of the $t\bar{t}$ system decay hadronically.

In the hadronic decay of the top, there is a $b$-jet and two light flavor jets, one of which is initiated by an up-type quark and one of which is initiated by a down-type quark.  If the down-type-initiated jet could be identified, then the maximal correlation of $\kappa=1.00$ would be maintained because the leading order matrix element for the down quark and lepton in top decays is the same.  Unfortunately, this is usually not possible.  One can consider charm tagging since charm quarks are present in half of the hadronic top decays.  It turns out that the charm tagging rate is not high enough for this to be better than the optimal hadronic method that we use. The required charm tagging rate is calculated in Appendix~\ref{app:ctagging}.\footnote{Another possibility would be incorporating measurements of jet charge, however, this seems challenging~\cite{Fraser:2018ieu,Kang:2023ptt}.}  In any case, in many studies of top spin correlations the softer of the two jets was used since one expects that down-type-initiated jet is more often the softer one.  This yields a spin analyzing power of $\kappa=0.50$. Using the $b$-jet is not ideal because its spin analyzing power is $\kappa=0.40$.  

\begin{figure} [tb]
  \begin{center}
  \includegraphics[width=0.7\textwidth]{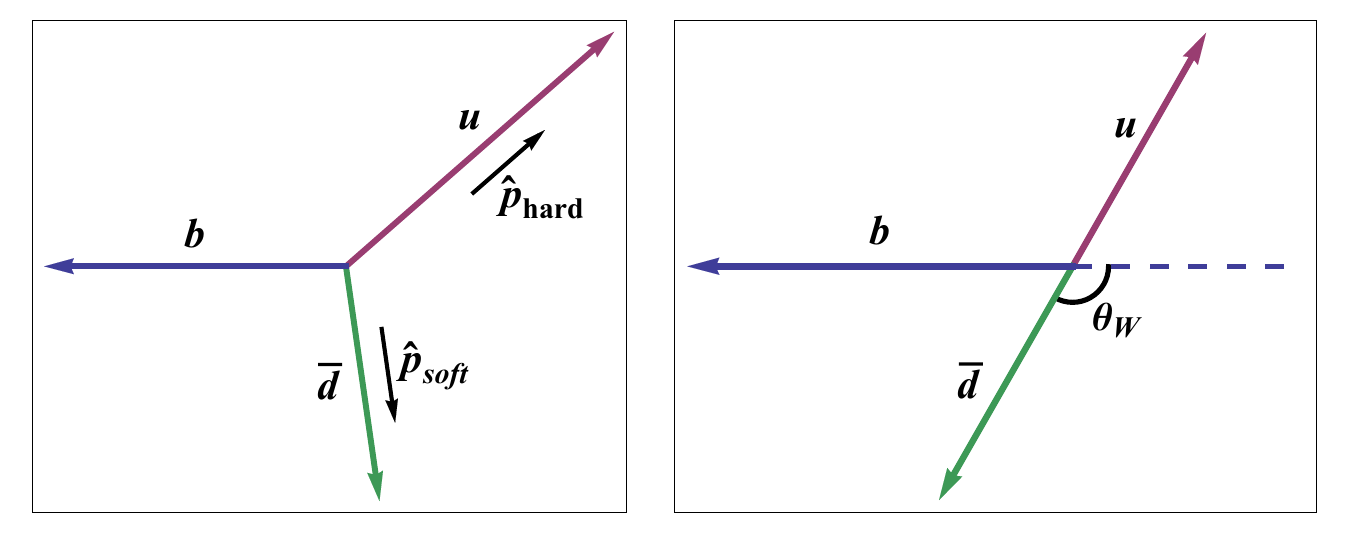}
  \caption{Illustration of the top decay system in the rest frame of the $t$ (left) and rest frame of the $W$ (right).  Between the down-type anti-quark and the up-type quark in the $t$ rest frame, the down-type anti-quark tends to be softer while the up-type quark tends to be harder.}
  \label{fig:wdecay_diagram}
  \end{center}
\end{figure}

The optimal spin analyzing power, assuming that one cannot distinguish the up-type-initiated and down-type-initiated jets, was calculated in Ref.~\cite{Tweedie:2014yda}.  They find an integrated value of $\kappa_{\rm opt}=0.64$ when one uses a weighted sum of the two jets whose four-vectors are labelled as $\vec{p}_{\rm soft}$ and $\vec{p}_{\rm hard}$.  The optimal hadronic value is given by using the four-vector $\vec{p}_{\rm opt}$ which is
\begin{equation} \label{eq:popt}
\vec{p}_{\mathrm{opt}} (\cos\theta_W)
= P_{d \to p_{\rm soft}}(\cos\theta_W) \, \hat{p}_{\mathrm{soft}} +
P_{d \to p_{\rm hard}}(\cos\theta_W) \, \hat{p}_{\mathrm{hard}},
\end{equation}
where $\theta_W$ is the angle between the momentum of the $d$-quark and the momentum axis of the $W$ in the rest frame of the $W$ (see Fig.~\ref{fig:wdecay_diagram}).  The function $P_{d \to p_{\rm soft}}(\cos\theta_W)$ is the probability that the $d$ quark is the softer jet and $P_{d \to p_{\rm hard}}(\cos\theta_W)$ is the probability that the $d$ quark is the harder jet.  These functions are given in Appendix~\ref{app:spin}.

The optimal direction for the hadronic decay of anti-top quark is defined likewise and the resulting spin analyzing power is $\kappa_{\rm opt} = -0.64$.  When extracting the components of the spin correlation matrix via Eq.~\eqref{eq:cij_individual} in the semi-leptonic channel one of the spin analyzing powers is given by the lepton and one is given by the optimal hadronic direction.

%%%%%%%%%%%%%
%%%%%%%%%%%%%

%%%%%%%%%%%%%%%%%%%%%%%%%%%%%%%%%%%%%%%%%%%%%%%%%%%%%%%%%%%%%%%%%%%%%%
\subsection{Entanglement in \texorpdfstring{$t\bar{t}$}{tt}}
\label{sec:entanglement_in_tt}

For a general bipartite quantum state,  the 15 values of $B^\cA_i$, $B^\cB_i$, and $C_{ij}$ need to be specified.  In the $t\bar{t}$ system at leading order, $B^\cA_i = 0$ and $B^\cB_i=0$ for all $i$, and $C_{ij} = C_{ji}$~\cite{Bernreuther:2015yna}.  Furthermore, in the helicity basis, where $1=\hat{r}$, $2=\hat{k}$, and $3=\hat{n}$, only $C_{12}$ is non-zero leading to a set of only 4 parameters: $C_{11}$, $C_{22}$, $C_{33}$, and $C_{12}$.

With only these parameters, a subset of the list of sufficient conditions generated by the PPT criterion can be shown to be a set of necessary conditions~\cite{Aguilar-Saavedra:2022uye}
\begin{subequations}
\label{eq:necessary1}
\begin{align}
|C_{11} + C_{22}| &> 1 + C_{33}, \label{eq:necessary1a} \\
|4 C_{12}^2 + (C_{11} - C_{22})^2|^{1/2} &> 1 - C_{33}.
\end{align}\end{subequations}
Instead of $\{ C_{11}, C_{22}, C_{33}, C_{12} \}$, one can use the three eigenvalues of the $C$ matrix
 $\{ C_1, C_2, C_3 \}$.  Using these Eq.~\eqref{eq:necessary1} becomes
\begin{subequations}
\label{eq:necessary2}
\begin{align}
|C_1 + C_2| &> 1 + C_3, \\
|C_1 - C_2| &> 1 - C_3.
\end{align}\end{subequations}
These conditions can be shown to be directly related to the concurrence
\begin{equation} \label{eq:concurrence}
\cC(\rho) = \left\{ \begin{array}{lr}
\frac{1}{2} \mathrm{max} (|C_1+C_2|-1-C_3,0), \quad\quad\quad
C_3\leq 0 \\ [2mm]
\frac{1}{2} \mathrm{max} (|C_1-C_2|-1+C_3,0), \quad\quad\quad
C_3\geq 0
\end{array}\right.
\end{equation}
where the necessary and sufficient condition for entanglement becomes the usual $\cC(\rho) > 0$. 

\begin{figure}  [tb]
  \begin{center}
  \includegraphics[width=0.55\textwidth]{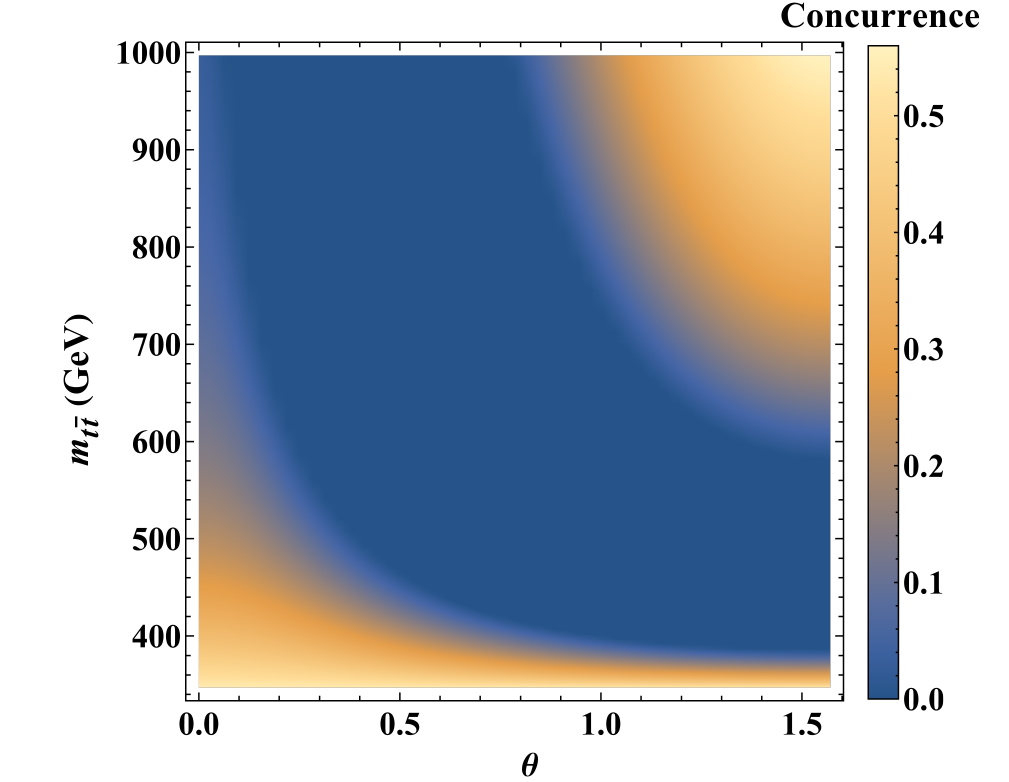}
  \caption{Concurrence $\cC(\rho)$ of the $t\bar{t}$ system at parton-level in the $\theta-m_{t\bar t}$ plane at $\sqrt{s}=13~{\rm TeV}$ with no phase space cuts. Entanglement is indicated by a value $\cC(\rho)>0$.} 
  \label{fig:concurrence_mtt_theta} 
  \end{center}
\end{figure}

The concurrence for $t\bar{t}$ is shown in Fig.~\ref{fig:concurrence_mtt_theta} as a function of phase space position $(\theta, m_{t\bar{t}})$ generated at parton-level at 13 TeV with no phase space cuts applied.
There is one region of sizable entanglement near threshold (due to like-helicity gluons producing a spin-singlet state) and a second region at high boost and large $\theta$ (due to unlike-helicity gluons producing a spin-triplet state).

%%%%%%%%%%%%%%%%%%
%%%%%%%%%%%%%%%

Since we have specified the final state to be $t\bar{t}$, for a given phase space region we can identify which case of Eq.~\eqref{eq:necessary1a} applies.  This can be done either semi-analytically through the spin production matrix in Eq.~\eqref{eq:productionSpinDensityMatrix} or numerically from Fig.~\ref{fig:concurrence_mtt_theta}.  Consider first the entangled region near threshold.  Here $C_{11} < 0$, $C_{22} < 0$, and $C_{33} < 0$, therefore by Eq.~\eqref{eq:concurrence} the inequality is
\begin{equation} \label{eq:ppt_threshold}
- C_{11} - C_{22} - C_{33} - 1 > 0.
\end{equation}
In the boosted region $C_{11} > 0$, $C_{22} > 0$, and $C_{33} < 0$, which leads to the inequality
\begin{equation} \label{eq:ppt_boosted}
C_{11} + C_{22} - C_{33} - 1 > 0.
\end{equation}
Converting these into operators, like in Eq.~\eqref{eq:sampleOpEnt}, we find
\begin{align} \label{eq:observable_d}
D &= \frac{1}{3} (C_{11} + C_{22} + C_{33}), 
\quad\quad\quad
&D < -\frac{1}{3}\quad 
{\rm for\  entanglement,} 
\\
\label{eq:observable_d3}
D_3 &= \frac{1}{3} (-C_{11} - C_{22} + C_{33}),
\quad\quad\quad
&D_3 < -\frac{1}{3}\quad 
{\rm for\  entanglement.} 
\end{align}
Through Eq.~\eqref{eq:observablesToAngles} both $D$ and $D_3$ can be related directly to measurements
\begin{align}
\frac{1}{\sigma} \frac{\dd \sigma}{\dd \cos{\theta^{\cA\cB}} } 
&= \frac{1}{2} (1+\kappa^\cA \kappa^\cB D \cos{\theta^{\cA\cB}}), \label{eq:disD} \\
\frac{1}{\sigma} \frac{\dd \sigma}{\dd \cos{\theta'^{\cA\cB}} } 
&= \frac{1}{2} (1+\kappa^\cA\kappa^\cB  D_3 \cos{\theta'^{\cA\cB}}),  \label{eq:disE}
\end{align}
where the angles are given by
\begin{align}
\cos \theta^{\cA\cB} &=
\sum_i \Omega_i^\cA \Omega_i^\cB,\\
\cos{\theta'^{\cA\cB}} &= \sum_{i,j}
\Omega_i^\cA P_{ij} \Omega_j^\cB.
\end{align}
The vector $\Omega_i^\cA$ is the normalized three-momentum of the decay product of the top in the top rest frame and the vector $\Omega_i^\cB$ is the normalized three-momentum of the decay product of the anti-top in the anti-top rest frame. In the fully leptonic channel these would be the anti-lepton and lepton.  In the semi-leptonic channel these would be anti-lepton or lepton and the optimal hadronic direction defined in Sec.~\ref{sec:spin}.  The matrix $P_{ij}$ is ${\rm diag}(-1,-1,1)$~\cite{Aguilar-Saavedra:2022uye}.

Extracting $D$ and $D_3$ via the asymmetry yields
\begin{align} 
\label{eq:d_direct}
D = \frac{4}{\kappa^\cA \kappa^\cB}
\left( A_{\cos{\theta^{\cA\cB}}} \right), \\
\label{eq:d3_direct}
D_3 = \frac{4}{\kappa^\cA \kappa^\cB}
\left( A_{\cos{\theta'^{\cA\cB}}} \right).
\end{align}
Measuring a quantity with a single observable was called the ``direct'' method in Ref.~\cite{Aguilar-Saavedra:2022uye}.  By contrast, measuring a quantity by first measuring each $C_{ij}$ value individually, then combining them, was called the ``individual'' method.  In the case of $(C_{11}+C_{22}+C_{33})/3$ using Eq.~\eqref{eq:d_direct} is the direct method while using Eq.~\eqref{eq:cij_individual} is the individual method.  Ref.~\cite{Aguilar-Saavedra:2022uye} argued that the direct method naively has slightly better sensitivity since there is only one uncertainty whereas for the individual method, multiple quantities are measured so their uncertainties are combined.  

From Eqs.~\eqref{eq:concurrence},~\eqref{eq:observable_d}, and~\eqref{eq:observable_d3} one sees that $\cC(\rho) = -3D-1$ in the threshold region and $\cC(\rho) = -3D_3-1$ in the boosted region.  $D$ is basis-independent because it is proportional to the trace of the spin correlation matrix $C$.  $D_3$ is basis-dependent and we use the helicity basis.  Experimentally, $D$ has been measured by CMS~\cite{CMS:2019nrx}.  They found a value of $-0.237 \pm 0.011$ without implementing an upper cut on $m_{t\bar{t}}$~\cite{Afik:2020onf}.  More recently, ATLAS measured $-0.547 \pm 0.02$ using an upper cut of $380~{\rm GeV}$ on $m_{t\bar{t}}$~\cite{ATLAS-CONF-2023-069}.

%%%%%%%%%%%%%%%%%%%%%%%%%%%%%%%%%%%%%%%%%%%%%%%%%%%%%%%%%%%%%%%%%%%%%%
\subsection{Bell Inequality Violation in \texorpdfstring{$t\bar{t}$}{tt}} \label{sec:bell_in_tt}

\begin{figure} [tb]
  \begin{center}
  \includegraphics[width=0.5\textwidth]{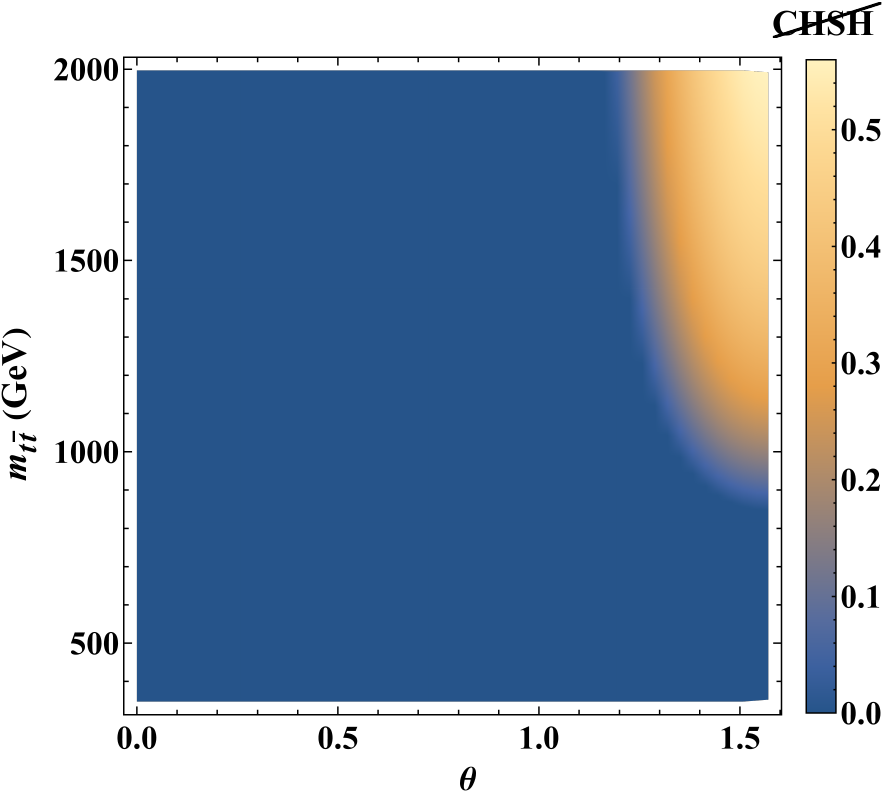}
  \caption{The CHSH violation $(B-\sqrt{2})$ of the $t\bar{t}$ system at parton-level in the $\theta-m_{t \bar t}$ plane at $\sqrt{s}=13~{\rm TeV}$ with no phase space cuts. 
 CHSH violation is indicated by a value $(B-\sqrt{2})>0$. Values of $(B-\sqrt{2})$ that are $<0$ are plotted as $0$.} 
  \label{fig:CHSH_mtt_theta}
  \end{center}
\end{figure}

To test Bell's inequality, we use the CHSH inequality, given in Eq.~\eqref{eq:CHSH}.  Using fixed axes in the CHSH inequality, in the helicity basis this corresponds to the operator\footnote{At high $p_T$ the helicity basis is known to result in large spin correlations~\cite{Mahlon:1995zn} while near threshold the fixed beam basis has larger spin correlations.  In the $t\bar{t}$ system due to the contributions from the $gg$ and $q\bar{q}$ initial states, the Bell inequality violation near threshold is very small.  For that reason we use the helicity basis and focus on the boosted region.  Eq.~\eqref{eq:CHSHineq} allows different choices for $C_{33}-C_{11}$ and we choose $C_{nn} - C_{rr}$ since it leads to the largest Bell inequality violation.}
\begin{equation} \label{eq:CHSHinHelicity}
B = C_{nn} - C_{rr},
\quad\quad\quad
B > \sqrt{2}\quad 
\text{ for Bell inequality violation.} 
\end{equation}
In Fig.~\ref{fig:CHSH_mtt_theta}, we show $B-\sqrt{2}$ in the phase space plane of $\theta - m_{t\bar{t}}$.  We see that Bell inequality violation is more appreciable at large $m_{t\bar{t}}$ and large $\theta$.

%%%%%%%%%%%%%%%%%%
%%%%%%%%%%%%%%%

We show how to construct the direct observable for $B=C_{nn} - C_{rr}$, following Ref.~\cite{Aguilar-Saavedra:2022uye}.  Consider the azimuthal angles $\phi^\cA$ and $\phi^\cB$.  The azimuthal angle $\phi$ is the angle around the $\hat{k}$ direction with the $\phi=0$ in the $\hat{n}- \hat{k}$ plane.
We construct
\begin{equation}
\phi_+ = \frac{\phi^\cA+\phi^\cB}{2},
\quad\quad\quad
\phi_- = \frac{\phi^\cA-\phi^\cB}{2}.
\end{equation}
From Eq.~\eqref{eq:observablesToAngles} one obtains
\begin{equation}\begin{aligned}
\frac{1}{\sigma}\frac{\dd \sigma}{\dd \phi_+ \dd \phi_-} 
= \frac{1}{2\pi^2} + \frac{\kappa^\cA \kappa^\cB}{32}
& \left(\frac{C_{nn}+C_{rr}}{2} \cos (2\phi_-) + \frac{C_{nn}-C_{rr}}{2} \cos(2\phi_+) \right. \\
& \left. \quad + \frac{C_{nr}+C_{rn}}{2} \sin(2\phi_+) + \frac{C_{rn}-C_{nr}}{2} \sin(2\phi_-) \right).
\end{aligned}\end{equation}
The term proportional to $C_{nn} - C_{rr}$ is the only term that is an even function with respect to $\phi_+$ so it can be extracted through the asymmetry
\begin{equation} \label{eq:b_direct}
% C_{nn}-C_{rr}=\frac{16}{ \pi\, \kappa^\cA \kappa^\cB} \, \frac{N^+_{\cos(2\phi_+)} - N^-_{\cos(2\phi_+)}}{N^+_{\cos(2\phi_+)} + N^-_{\cos(2\phi_+)}}.
B=C_{nn}-C_{rr}=\frac{16}{ \pi\, \kappa^\cA \kappa^\cB} 
\left( A_{\cos(2\phi_+)} \right).
\end{equation}
Alternatively, we derive the full functional form by integrating out $\phi_-$ and making a change of variable.  Defining $\phi_P = \frac{\pi}{2}-\big|\frac{\pi}{2}-|\pi-\phi_+|\big|$, after integrating out $\phi_-$, the distribution is then
\begin{equation} \label{eq:CHSHbyFit}
\frac{1}{\sigma}\frac{\dd \sigma}{\dd \phi_P} = \frac{2}{\pi} + \frac{\pi}{16} \kappa^\cA \kappa^\cB \, (C_{nn}-C_{rr}) \cos (2\phi_P).
\end{equation}
Since we have derived the full functional form in Eq.~\eqref{eq:CHSHbyFit}, the value of $B=C_{nn}-C_{rr}$ can also be extracted by a fit.

%%%%%%%%%%%%%%%%%%%%%%%%%%%%%%%%%%%%%%%%%%%%%%%%%%%%%%%%%%%%%%%%%%%%%%
%%%%%%%%%%%%%%%%%%%%%%%%%%%%%%%%%%%%%%%%%%%%%%%%%%%%%%%%%%%%%%%%%%%%%%
\section{Results at the LHC}
\label{sec:lhc}

%%%%%%%%%%%%%%%%%%%%%%%%%%%%%%%%%%%%%%%%%%%%%%%%%%%%%%%%%%%%%%%%%%%%%%
\subsection{Sketch of Expected Results}
\label{sec:sketch}

Consider an observable $\cO$ that is sensitive to the presence of entanglement.  One example would be the observable in Eq.~\eqref{eq:sampleOpEnt}.
A useful observable will have a large difference between the measured value $\cO_{\rm entangled}$ for an entangled state and the predicted value $\cO_{\rm null}$ for a separable state with no entanglement.  

Let the measured value of the observable be $\cO_{\rm entangled} \pm \delta \cO$ (corresponding to one standard deviation).  The significance can be approximated by
\begin{equation}
\text{significance} \approx 
\frac{\cO_{\rm entangled} - \cO_{\rm null}}{\delta \cO}.
\end{equation}
The sensitivity of the observable can be increased either by reducing the uncertainty $\delta \cO$ (for example, by collecting more data) or by choosing a quantum state with a larger expected value of $\cO_{\rm entangled}$ (for example, through phase space cuts).

%%%%%%%%%%%%%%%%%%%%%%%%%%%%%%%%%%%%%%%%%
\paragraph{Reducing the uncertainty:}

The leptonic decay channels of $W \to \ell \nu\ (\ell = e, \mu)$ have a branching fraction  $\text{BR}(W \to \ell \nu) = 0.21$~\cite{ParticleDataGroup:2022pth}.  The branching fraction of $t\bar{t}$ into the fully leptonic channel is thus 
\begin{equation}
\text{BR}(t\bar{t} \to \ell \ell)
= 0.0455.
\end{equation}
There the complete final state consists of $\ell^- \nu \bar{b} \ell^+ \bar{\nu} b$, but we write it as $\ell \ell$ for simplicity.  The hadronic branching fraction of the $W$ decay is $\text{BR}(W \to \text{hadrons}) = 0.67$~\cite{ParticleDataGroup:2022pth}, so the branching fraction of $t\bar{t}$ into the semi-leptonic channel is
\begin{equation}
\text{BR}(t\bar{t} \to \ell j)
= 0.2877,
\end{equation}
which is about a factor of 6 larger than the fully leptonic channel.  Again, we've written the final state as $\ell j$ which represents either $\ell^- \nu \bar{b} q \bar{q}' b$ or $q \bar{q}' \bar{b} \ell^+ \bar{\nu} b$.

Assuming that the uncertainty on $\cO$ is statistics dominated, the uncertainty in the channel $ij$ will scale as $1/\sqrt{{\rm BR}(\bar{t}t \to ij)}$.  Relative to the fully leptonic channel, we expect that the uncertainty on $\cO$ in the semi-leptonic channel is decreased by a factor of $\sqrt{{\rm BR}(\bar{t}t \to \ell\ell) / {\rm BR}(\bar{t}t \to \ell j) }$ or a gain of a factor of $2.5$.

%%%%%%%%%%%%%%%%%%%%%%%%%%%%%%%%%%%%%%%%%
\paragraph{Naive Expectation:}

The correlation between the polarization of the top (or anti-top) and one of its decay products $i$ is given by the spin analyzing power $\kappa_i$, as discussed in Sec.~\ref{sec:spin} and Appendix \ref{app:spin}.  The spin analyzing power of the anti-lepton (or lepton) in the top (or anti-top) decay is $|\kappa_\ell| = 1$; it is maximally correlated with the polarization of the top (or anti-top).  For the hadronic decay of the top (or anti-top) the spin analyzing power is smaller and it is $|\kappa_q| = 0.64$.  In the semi-leptonic channel, the leptonically-decaying top (or anti-top) uses the anti-lepton (or lepton) as a proxy for the polarization and the hadronically-decaying top (or anti-top) uses the jets as a proxy for the polarization.  The observable for semi-leptonic channel has a different pre-factor (see Eq.~\eqref{eq:cij_individual}) than the fully leptonic channel which scales the uncertainty by a factor of $| (\kappa_\ell \kappa_\ell)/(\kappa_\ell \kappa_q) | = 1/0.64$.

Combining both of the previous effects, we expect that the relative significance between the decay channel $t\bar{t} \to ab$ and $t\bar{t} \to cd$ is given by
\begin{equation} \label{eq:rescaling}
\frac{\text{significance}\;(t\bar{t} \to ab)}{\text{significance}\;(t\bar{t} \to cd)}
= \frac{\kappa_a \kappa_b}{\kappa_c \kappa_d}
\sqrt{\frac{\text{BR}(t\bar{t} \to ab)}{\text{BR}(t\bar{t} \to cd)}}.
\end{equation}
Comparing the semi-leptonic to the leptonic we have
\begin{equation}\label{eq:improvement}
\frac{\text{significance}\;(t\bar{t} \to \ell q)}{\text{significance}\;(t\bar{t} \to \ell \ell)}
= 0.64
\sqrt{\frac{0.2877}{0.0455}}
= 1.60. 
\end{equation}
We naively expect an improvement of $60\%$. There will also be a further  improvement in the reconstruction efficiency of the semi-leptonic channel because there is a single neutrino as opposed to the fully leptonic channel which has two neutrinos, as we will exploit in our full analysis of the semi-leptonic channel.
Following the scaling as in  Eq.~\eqref{eq:improvement}, the fully hadronic channel is expected to gain 29\% over the fully leptonic channel. Given the challenges for the signal identification and background suppression for the fully hadronic channel, we leave this to a future study.  

%%%%%%%%%%%%%%%%%%%%%%%%%%%%%%%%%%%%%%%%%%%%%%%%%%%%%%%%%%%%%%%%%%%%%%
\subsection{Simulation}
\label{sec:simulation}

We perform our analyses in two stages.  The first is ``parton-level'', where events are generated without parton shower or hadronization.  The uncertainty for parton-level events is always just statistical from the number of events.  We further carry out a ``detector-level'' (or ``reconstructed'') study, which includes parton showering, hadronization, detector simulation, and event reconstruction.  Parameters extracted from the detector-level analysis are always corrected using parametric fitting (see Sec.~\ref{sec:unfolding}) and the uncertainties include the impact of the parametric fitting.  In the few instances where detector-level results are shown without parametric fitting it will be noted explicitly. 

All events are generated with {\tt Madgraph 5}~\cite{Alwall:2011uj} at $\sqrt{s}=13~{\rm TeV}$ using the {\tt NNPDF 2.3} parton distribution function~\cite{Ball:2013hta}.  Three samples are generated: a $t\bar{t}$ sample that decays through the fully leptonic channel and two $t\bar{t}$ samples that decay through the semi-leptonic channel.  In all samples we generate $pp \to t\bar{t}$ at leading order and then the events are decayed using {\tt Madspin}~\cite{Artoisenet:2012st}.  We apply a flat $k$-factor of 1.8 to account for the QCD correction to the total cross section~\cite{Czakon:2011xx}.

The leptonic sample includes the decays $t\bar{t} \to (b \ell^+ \nu_\ell)(\bar{b} \ell^- \bar{\nu}_\ell)$ where $\ell = e, \mu$.  It is generated with no phase space cuts and only at parton-level.  The semi-leptonic samples include both $t\bar{t} \to (b \ell^+ \nu_\ell)(\bar{b} q \bar{q}')$ and $t\bar{t} \to (b q \bar{q}')(\bar{b} \ell^- \bar{\nu}_\ell)$ where $q, q'$ are light flavor quarks. The partonic final states are then showered and hadronized with {\tt Pythia 8}~\cite{Sjostrand:2014zea} and go through the detector simulation {\tt Delphes 3}~\cite{deFavereau:2013fsa}.  We have two semi-leptonic samples in two kinematic regions according to the invariant mass of the $t\bar t$ system.

%%%%%%%%%%%%%%%%%%%%%%%%%%%%%%%%%%%%%%%%%
\paragraph{Resolved sample:}
We first start with the semi-leptonic sample with no additional phase space cuts, which we call the ``resolved sample.'' 
The event selection used is
\begin{subequations} \label{eq:objSelection}
\begin{align} 
p_T(j) &> 25~{\rm GeV},\quad\quad
|\eta(j)| < 2.5, \label{eq:objslct1}   \\
p_T(\ell) &> 25~{\rm GeV},\quad\quad
|\eta(\ell)| < 2.5, \\
\slashed{E}_T &> 30~{\rm GeV}.
\end{align}
\end{subequations}
Jets are clustered with the anti-$k_T$ algorithm with a separation $\Delta R = \sqrt{\Delta\phi^2 + \Delta\eta^2} =0.5$~\cite{Cacciari:2008gp}.  
This is approximately the event selection corresponding to a single lepton trigger~\cite{ATLAS:2020aln,CMS:2023qyl}. 

To compute the spin correlation matrix,  it is necessary to fully reconstruct the final state kinematics.  This requires identifying two $b$-jets, estimating the four-vector of the neutrino (or anti-neutrino), and assigning each $b$-jet to either the leptonic pair or the jet pair.  We employ a modified version~\cite{compare_reconstruct} of the pseudo-top algorithm~\cite{ATLAS_reconstruct,Collaboration:2267573}.  If the event contains only one $b$-tagged jet, the hardest jet from the non-$b$-tagged jet is assumed to be the second $b$-jet.  The neutrino (or anti-neutrino) four-vector is determined from the two components of the missing transverse energy vector and from solving the on-shell condition of the neutrino and the on-shell condition of the leptonically-decaying $W$ boson. 

\begin{figure} [tb]
  \begin{center}
  \includegraphics[width=0.7\textwidth]{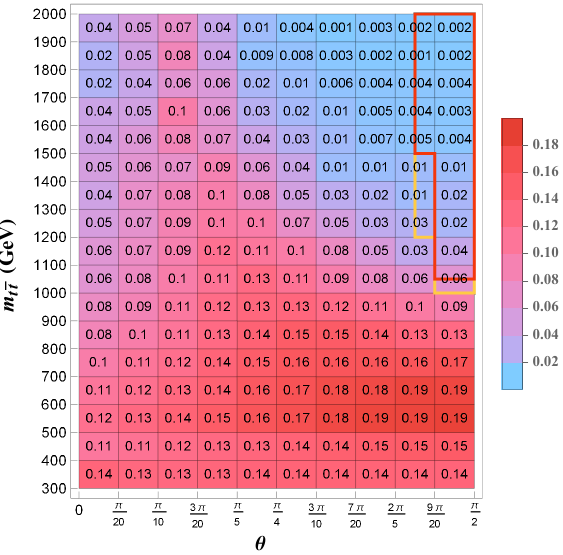}
  \caption{Reconstruction efficiency in the $\theta -m_{t\bar{t}}$ plane for the resolved semi-leptonic sample at $\sqrt{s}=13~{\rm TeV}$.  The weak (orange line) and strong (red line) regions correspond to signal regions in Sec.~\ref{sec:regions}.}
  \label{fig:reconstruction_efficiency_resolved}
  \end{center}
\end{figure}

The resulting reconstruction efficiency is defined as the number of events that are successfully reconstructed compared to the total events generated.  The differential reconstruction efficiency is shown in Fig.~\ref{fig:reconstruction_efficiency_resolved}. We find that the reconstruction efficiency peaks around 19\% near threshold and decreases as the invariant mass of the system, and consequently the boost of the top and anti-top, increases. 

%%%%%%%%%%%%%%%%%%
%%%%%%%%%%%%%%%

%%%%%%%%%%%%%%%%%%%%%%%%%%%%%%%%%%%%%%%%%
\paragraph{Boosted sample:}

The other semi-leptonic sample is generated with 
\begin{equation}
m_{t\bar{t}} > 800~{\rm GeV},
\end{equation}
at parton-level and we call it the ``boosted sample.''   This corresponds to a boost factor $\gamma>2.3$ in the center-of-mass frame  for a fast-moving top quark.

We first cluster events with the anti-$k_T$ algorithm into jets with $\Delta R_{\rm sub} = 0.2$ and apply the event selection from Eq.~\eqref{eq:objSelection}.  In the boosted sample we will call these subjets, even though they are clustered from the full event.  We then recluster the event into ``fat jets''
\begin{equation}
\Delta R_{\rm fat} = 1.5,
\quad\quad\quad
|\eta|<2.5,
\quad\quad\quad
p_T>300~{\rm GeV}.
\end{equation}
A single fat jet $J$ is matched to three subjets $j_{\rm sub}$ by selecting the three highest $p_T$ subjets that satisfy
\begin{equation}
\Delta R ( J , j_{\rm sub} ) < \Delta R_{\rm fat}.
\end{equation}
The three matched subjets are required to constitute most of the transverse momentum of the fatjet
\begin{equation}
\frac{p_T(j_{{\rm sub}_1} + j_{{\rm sub}_2} + j_{{\rm sub}_3})}{p_T(J)} > 0.9.
\end{equation}
The hadronic top is then taken to be the four-vector sum of the three subjets $p_{\rm top} = p_{j_{{\rm sub}_1}} + p_{j_{{\rm sub}_2}} + p_{j_{{\rm sub}_3}}$.  This procedure approximately corresponds to the fat jets and corresponding subjets that would result from the trimming procedure~\cite{Krohn:2009th}.

One of the three matched subjets is expected to be $b$-tagged. If none are $b$-tagged, the highest $p_T$ of the matched subjets is assumed to be the $b$-jet.  Finally, the mass of the hadronic top $m = \sqrt{p_{\rm top}^2}$ is required to be close to the 175 GeV, and we choose it in the range $(150~{\rm GeV}, 225~{\rm GeV})$.

\begin{figure} [tb]
  \begin{center}
  \includegraphics[width=0.7\textwidth]{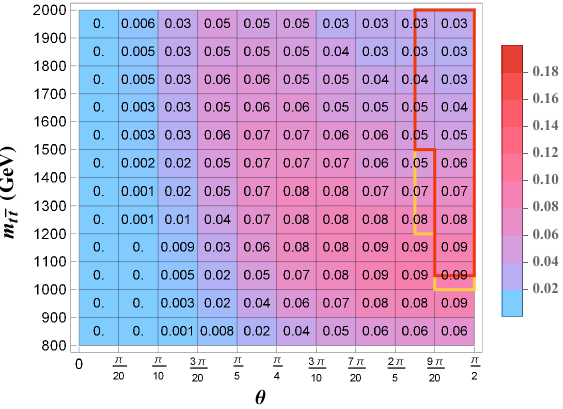}
  \caption{Reconstruction efficiency in the $\theta- m_{t\bar{t}}$ plane for the boosted semi-leptonic sample generated at $\sqrt{s}=13~{\rm TeV}$.  The weak (orange line) and strong (red line) regions correspond to signal regions in Sec.~\ref{sec:regions}.  The reconstruction efficiency is higher in the boosted sample than in the resolved sample for these two signal regions.}
  \label{fig:reconstruction_efficiency_boosted}
  \end{center}
\end{figure}

The differential reconstruction efficiency for the boosted selection is shown in Fig.~\ref{fig:reconstruction_efficiency_boosted}.  For moderately boosted tops where $m_{t\bar{t}} \lesssim 1000~{\rm GeV}$ the resolved sample has a higher reconstruction efficiency.  For tops with $m_{t\bar{t}} \gtrsim 1000~{\rm GeV}$ the reconstruction efficiency of the boosted selection is better by more than an order of magnitude. We find that even for $m_{t\bar{t}} \approx 2~{\rm TeV}$ the reconstruction efficiency remains above 3\% in the signal regions where $\theta \approx \pi/2$.

%%%%%%%%%%%%%%%%%%
%%%%%%%%%%%%%%%

%%%%%%%%%%%%%%%%%%%%%%%%%%%%%%%%%%%%%%%%%%%%%%%%%%%%%%%%%%%%%%%%%%%%%%
\subsection{Unfolding and Parametric Fitting}
\label{sec:unfolding}

While the events selection cuts in Eq.~\eqref{eq:objSelection} are very minimal in terms of the event identification, they have a sizable impact on the angular distributions that are used to extract the spin correlation coefficients.

\begin{figure} [tb]
  \begin{center}
  \includegraphics[width=0.65\textwidth]{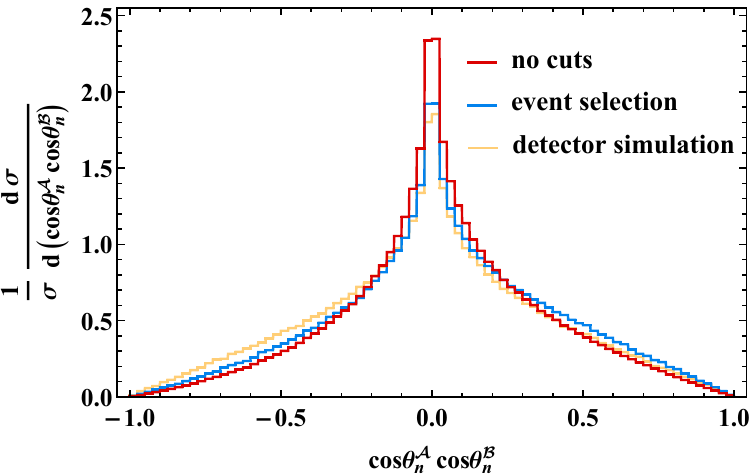}
  \caption{Parton-level differential distribution of $\cos\theta^\cA_n \cos\theta^\cB_n$ at $\sqrt{s}=13~{\rm TeV}$ with no cuts applied (red) and with event selection from Eq.~\eqref{eq:objSelection} (blue) and with detector simulation (yellow). } 
  \label{fig:distribution_cosAcosB}
  \end{center}
\end{figure}

For example, consider the distribution of $\cos\theta_n^\cA \cos\theta_n^\cB$ which, by Eq.~\eqref{eq:cosAcosB}, can be used to measure $C_{nn}$. In Fig.~\ref{fig:distribution_cosAcosB}, the red line shows the differential distribution of $\cos\theta_n^\cA \cos\theta_n^\cB$ with no phase space cuts.  This is the distribution that would be used to extract the value of $C_{nn}$.  The effect of the event selection is shown by the blue line.  The selection distorts the distribution which invalidates the parameter estimation.  The yellow line shows the effects of the detector simulation which further alters the distribution.  In order to measure spin correlations accurately, it is necessary  to restore distributions to their inclusive shapes.

%%%%%%%%%%%%%%%%%%
%%%%%%%%%%%%%%%

Let $\vec{x}_{\rm truth}$ be the data if it could be measured with no detector effects or phase space cuts, and  $\vec{x}_{\rm detected}$ be the measured data.  We call the effect of the detector and cuts ``folding''
\begin{equation}
\vec{x}_{\rm truth}
\xrightarrow{\text{folding}} 
\vec{x}_{\rm detected} = R \cdot \vec{x}_{\rm truth},
\end{equation}
where the matrix $R$ is the response matrix.

Unfolding is the procedure that attempts to undo both detector effects and phase space cuts via $\vec{x}_{\rm truth} = R^{-1} \cdot \vec{x}_{\rm detected}$.  Generally, this is an ill-defined inversion problem which means that algorithm and regularization choices are required to obtain a result.  These choices are actually very important in the case of entanglement and Bell inequality violation because the experimental sensitivity is entirely driven by the obtainable uncertainty on spin correlation measurements.  Ideally the unfolding procedure itself would not substantially increase the uncertainty.

Let the uncertainty from statistics only be $\Delta _{\rm stat}$ and let the uncertainty after detector effects and unfolding be $\Delta_{\rm tot}$, such that for a given measurement the increase from statistics only is a factor of  $\Delta_{\rm tot}/\Delta_{\rm stat}$.  In Ref.~\cite{Severi:2021cnj} the increase is a factor of $1.46-1.53$ while in Ref.~\cite{Dong:2023xiw} the factor is $0.88$ (meaning that the final uncertainty is smaller than the statistics only uncertainty).

Past studies on this topic, including Refs.~\cite{Fabbrichesi:2021npl,Severi:2021cnj,Dong:2023xiw}, have used either the Iterative Bayesian (IB) method~\cite{dagostini2010improved} or the Singular Value Decomposition (SVD) method~\cite{Hocker:1995kb} implemented in either the {\tt RooUnfold} package~\cite{Adye:2011gm} or {\tt TSVDUnfold} package~\cite{Hocker:1995kb}.  For both of these methods one needs to choose both the number of bins to use in the unfolding and a parameter related to regularization.  In previous studies it was stated that the resulting uncertainty of spin correlation measurements was stable with respect to different choices. 

By contrast we find that variations to these parameters can change the resulting uncertainty by up to 75\%.  As there are many fewer events in the detected sample compared to the truth sample, some level of instability is expected.  These variations are shown in detail in Appendix~\ref{app:unfolding} along with results from an alternative unfolding method called the One-at-a-time Strict Bound method (OSB)~\cite{Stanley_2022}.

In our work, we apply the more common procedure used by the LHC experiments of parametric fitting.  While unfolding is typically applied at the level of distributions, parametric fitting is applied to the parameter estimation. 
Consider a parameter $\Theta$, then schematically parametric fitting can be described as
\begin{equation}
\vec{x}_{\rm truth}(\Theta)
\xrightarrow{\text{folding}} 
\vec{x}_{\rm predicted}(\Theta) = R \cdot \vec{x}_{\rm truth}(\Theta).
\end{equation}
The data $\vec{x}_{\rm detected}$ is fit to $\vec{x}_{\rm predicted}(\Theta)$ to extract the value of $\Theta$.  Here there is no need to invert the response matrix and therefore it is not dependent on a regularization parameter.  We find this method to be more stable and more intuitive than unfolding.  The uncertainty on the parameter $\Theta$ can be calculated by performing pseudo-experiments.  In our work, we carry out 1000 pseudo-experiments.  More details are presented in Appendix~\ref{app:unfolding}.

%%%%%%%%%%%%%%%%%%%%%%%%%%%%%%%%%%%%%%%%%%%%%%%%%%%%%%%%%%%%%%%%%%%%%%
\subsection{Signal Regions}
\label{sec:regions}

\begin{figure} [tb]
  \begin{center}
  \includegraphics[width=0.5\textwidth]{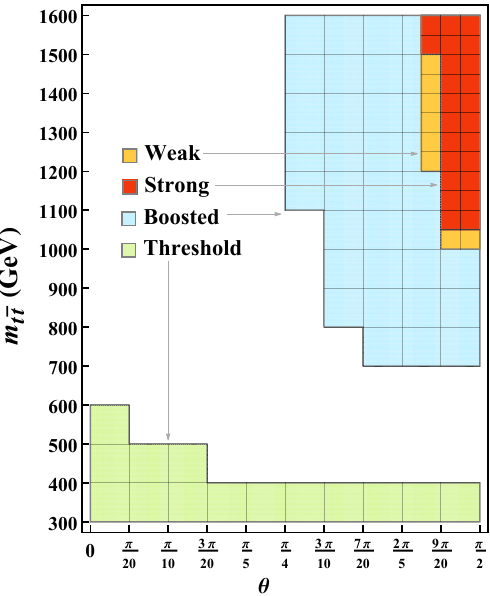}
  \caption{Signal regions in the $\theta- m_{t\bar{t}}$ plane. The regions for entanglement are: threshold (green) and boosted (blue).  The regions for Bell inequality violation are: weak (orange) and strong (red, overlap in orange).}
  \label{fig:signal_regions}
  \end{center}
\end{figure}

From Figs.~\ref{fig:concurrence_mtt_theta} and~\ref{fig:CHSH_mtt_theta}, it is clear that the size of entanglement and of Bell inequality violation differs over phase space.  To maximize the observable signals we specify four signal regions.  These are shown graphically in Fig.~\ref{fig:signal_regions} as non-rectangular cuts in the $\theta - m_{t\bar{t}}$ plane.

%%%%%%%%%%%%%%%%%%
%%%%%%%%%%%%%%%

The ``threshold'' region (the green region in Fig.~\ref{fig:signal_regions}) selects events that are very close to threshold.  There is an additional cut in this region to further enhance the significance, which is requiring the velocity of the $t\bar{t}$ system in the lab frame, $\beta = p_{t\bar t}/m_{t\bar t}$, to satisfy 
\begin{equation}
|\beta|\leq 0.9,
\end{equation}
as proposed by Ref.~\cite{Aguilar-Saavedra:2022uye}.  In this region, the $t\bar{t}$ pair is primarily produced in a spin singlet state from gluon fusion~\cite{Mahlon:2010gw}.  The $t\bar{t}$ cross section is also largest near threshold.  These facts together make this region ideal for detecting entanglement.

The ``boosted'' region (the blue region in Fig.~\ref{fig:signal_regions}) selects events where the top and anti-top are moderately boosted and the angle $\theta$ is sizable.  This region corresponds to the other entangled region from Fig.~\ref{fig:concurrence_mtt_theta}.  At high $p_T$, the $t\bar{t}$ pair is primarily produced in a spin triplet state from incoming gluons, however due to the falling cross section at larger $m_{t\bar{t}}$ we expected lower detection significance compared to the threshold region.

The ``weak'' region (the orange region in Fig.~\ref{fig:signal_regions}) selects events at larger $m_{t\bar{t}}$ and larger $\theta$.  From Fig.~\ref{fig:CHSH_mtt_theta}, it can be seen that unlike for entanglement, Bell inequality violation is only observable for large $m_{t\bar{t}}$. 

Finally, the ``strong'' region (the red region in Fig.~\ref{fig:signal_regions}) is even more restrictive on $m_{t\bar{t}}$ and $\theta$.  While the strong region is expected to more effectively isolate the phase space with Bell inequality violation, there are fewer events with the more restrictive cuts.  We include this region in addition to the weak region because a priori we do not know which region will have more sensitivity.
Note that the weak region is a subset of the boosted region and the strong region is a subset of both the boosted region and the weak region.

%%%%%%%%%%%%%%%%%%%%%%%%%%%%%%%%%%%%%%%%%%%%%%%%%%%%%%%%%%%%%%%%%%%%%%
\subsection{Entanglement Results}
\label{sec:results_entanglement}

Before presenting results on entanglement, in Table~\ref{tab:Cij_threshold} we show the measured values of the elements of the spin correlation matrix $C_{ij}$ in the helicity basis in the threshold region.  The values of $C_{ij}$ are measured using Eq.~\eqref{eq:cij_individual}.  Parton-level results contain no event selection and detector-level results are fully corrected.

The uncertainties on parton-level results are purely statistical while the uncertainties on detector-level results are larger because they include additional sources of uncertainty from the detector simulation and from the parametric fitting.  The uncertainties are different for different entries of the $C_{ij}$ matrix because each distribution gets distorted by detector effects in different ways.  The distribution itself also impacts the resulting uncertainty.  For the entries of the spin correlation matrix parametric fitting increases the uncertainties by a factor of $2 -4$. The outcomes, however, are quite stable and robust.

%%%%%%%%%%%%%%%%%%%%%%%%%%
\begin{table} [tb]
\begin{subtable}{1\textwidth}
\sisetup{table-format=-1.3}
\centering
\begin{tabular}{c | D{,}{\,\pm\,}{-1} D{,}{\,\pm\,}{-1} D{,}{\,\pm\,}{-1}} 
\hline
 ~parton-level~   & n & r & k \\ [0.5ex] 
 \hline
 $n$ & -0.500, 0.006 & 0.000, 0.006 & 0.000, 0.006 \\
 $r$ & -0.004, 0.006 & -0.361, 0.006 & -0.010, 0.006 \\
 $k$ & -0.006, 0.006 & -0.004, 0.006 & -0.656, 0.006 \\
\hline
\end{tabular}
\end{subtable}
%%%%%%%%%%%
%%%%%%%%%%%
%
\begin{subtable}{1\textwidth}
\sisetup{table-format=-1.3}
\centering
\begin{tabular}{c | D{,}{\,\pm\,}{-1} D{,}{\,\pm\,}{-1} D{,}{\,\pm\,}{-1}} 
   detector-level  & n & r & k \\ [0.5ex] 
 \hline
 $n$ & -0.510,0.012 & 0.000,0.023 & 0.001,0.019 \\
 $r$ & 0.001,0.022 & -0.359,0.023 & 0.000,0.030 \\
 $k$ & -0.005,0.019 & 0.000,0.026 & -0.655,0.020 
\end{tabular}
\end{subtable}
\caption{The spin correlation matrix $C_{ij}$ at parton-level (top) and at detector-level (bottom) in the threshold region generated at $\sqrt{s} = 13~{\rm TeV}$ with $\mathcal{L} = 139~{\rm fb}^{-1}$.} 
\label{tab:Cij_threshold}
\end{table}
%%%%%%%%%%%%%%%%%%%%%%%%%%

Results for entanglement are given by two times the concurrence $2\mathcal{C}(\rho)$, where the concurrence is given by Eq.~\eqref{eq:concurrence}.  Entanglement is indicated by $2\mathcal{C}(\rho)>0$.
The factor of two is included for easier comparison with other studies~\cite{Severi:2021cnj,Aguilar-Saavedra:2022uye}.  Results at parton-level are shown in Table~\ref{tab:concurrence_results} (top).
The uncertainty is purely statistical taking the number of events as $\epsilon N_{\rm parton}$ where $\epsilon$ is the average reconstruction efficiency for that signal region and $N_{\rm parton} = k \times \mathcal{L} \times \sigma_{\rm LO}$, where the $k$-factor is 1.8 and the luminosity for the existing LHC data is $139~{\rm fb}^{-1}$.  The individual results are calculated from Eq.~\eqref{eq:cij_individual} and the direct results are calculated from Eqs.~\eqref{eq:d_direct} and~\eqref{eq:d3_direct}.
Since all results are well above a significance of $5\sigma$, we show the precision which is given by $\Delta \cC(\rho) / \cC(\rho)$.

Comparing the threshold and boosted signal regions, we see that while the boosted region has a larger concurrence, the threshold region has about an order of magnitude of more events, yielding an uncertainty about 3 times smaller.  Furthermore, the direct method reduces the uncertainty on the parton-level results by about $20\%$ which is consistent with Ref.~\cite{Aguilar-Saavedra:2022uye}.

Entanglement results at detector-level after parametric fitting are shown in Table~\ref{tab:concurrence_results} (bottom).  The value of $N_{\rm detected}$ accounts for detector efficiencies.  The central value of $2\mathcal{C}(\rho)$ does not change relative to the parton-level result which is expected.  The uncertainty, however, is larger than the statistics-only result by roughly a factor of $3$ for the individual method and a factor of $2$ for the direct method.

\begin{figure} [tb]
  \begin{center}
  \includegraphics[width=0.47\textwidth]{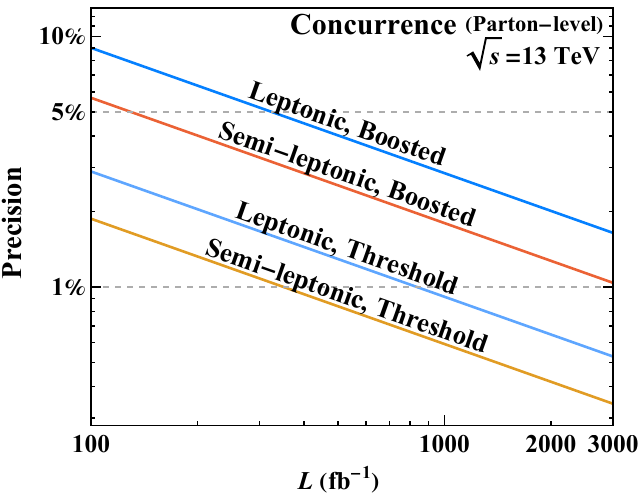}
  \quad\quad
  \includegraphics[width=0.47\textwidth]{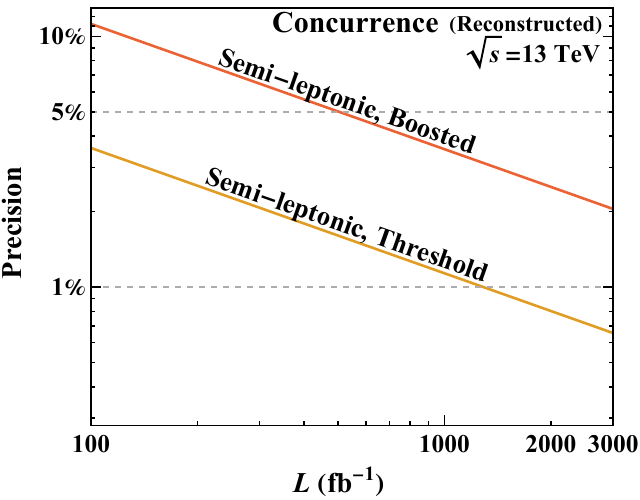}
  \caption{Expected precision of entanglement detection as a function of the integrated luminosity at the $13~{\rm TeV}$ LHC at parton-level (left) and after detector simulation, reconstruction, and parametric fitting (right).}
\label{fig:concurrence_results}
  \end{center}
\end{figure}

The precision as a function of luminosity is shown in Fig.~\ref{fig:concurrence_results} (left) at parton-level.  With only statistical errors, the parton-level result predicts that a $1\%$ precision can be achieved with around $300~{\rm fb}^{-1}$, corresponding to the end of LHC Run-3.  The results from the fully leptonic channel are also shown for comparison.  This channel is calculated at parton-level using the same efficiency that was calculated in the semi-leptonic sample.\footnote{The actual efficiency for the leptonic channel~\cite{Severi:2021cnj} is expected to be lower than in the semi-leptonic channel.}  Our calculation from Sec.~\ref{sec:sketch} predicted an improvement of $60\%$ which is what the parton-level result also finds.  Our leptonic result is consistent with Ref.~\cite{Aguilar-Saavedra:2022uye} (see Appendix~\ref{app:comparison} for a full comparison).

Figure~\ref{fig:concurrence_results} (right) shows the precision as a function of luminosity with the detector simulation. Including the detector effects increases the data required to reach $1\%$ precision to roughly $1200~{\rm fb}^{-1}$.  Even with the current LHC dataset, a detection of $5\sigma$ is still easily obtainable.

%%%%%%%%%%%%%%%%%%%%%%
\begin{table} [tb]
\centering
\begin{tabular}{||c||c c c c||c||} 
 \hline
\hspace{0.15em} \multirow{2}{5.3em}{Parton-level} \hspace{0.15em} & \multirow{2}{4em}{Efficiency} & $\epsilon \,N_{\mathrm{parton}}$ &\multicolumn{2}{c||}{\hspace{0.1em} $2\cC(\rho)$}  & \multirow{2}{4em}{Precision} \\ 
&  & $(139\, \mathrm{fb}^{-1})$ & (Individual) & (Direct) &  \\
 \hline
 Threshold & 0.16 & $1.26 \times 10^6$ & $0.518 \pm 0.010$ & $0.522 \pm 0.008$ & $1.6\%$ \\
 \hline
 Boosted & 0.13 & $1.15 \times 10^5$ & $0.576 \pm 0.032 $ & $0.566 \pm 0.027$ & $4.8\%$ \\ 
 \hline
\end{tabular}
\\[1mm]
\begin{tabular}{||c||c c c||c||} 
 \hline
\multirow{2}{6.3em}{Reconstructed}  & \hspace{2.6em}
$N_{\mathrm{detected}}$ \hspace{2.6em} & \multicolumn{2}{c||}{\hspace{0.1em} $2\cC(\rho)$} & \multirow{2}{4em}{Precision} \\
  & 
$(139\, \mathrm{fb}^{-1})$  & (Individual) &(Direct) & \\ 
 \hline
 Threshold & $1.26 \times 10^6$ & $0.523 \pm 0.033$ & $0.522 \pm 0.016$ & $3.0\%$ \\
 \hline
 Boosted & $1.15 \times 10^5$ & $0.549 \pm 0.084 $ & $0.552 \pm 0.052$ & $9.5\%$ \\ 
 \hline
\end{tabular}
\caption{Measurements of $2 \cC(\rho)$ generated at $\sqrt{s} = 13~{\rm TeV}$ and $\mathcal{L}=139~{\rm fb}^{-1}$ at parton-level (top) and after detector simulation, reconstruction, and parametric fitting (bottom).  Entanglement is indicated by $2 \cC(\rho)>0$.  The efficiency indicated is the average over the specified signal region.  The precision uses the direct measurement both at parton-level and at reconstruction-level.} 
\label{tab:concurrence_results}
\end{table}
%%%%%%%%%%%%%%%%%%%%%%

%%%%%%%%%%%%%%%%%%
%%%%%%%%%%%%%%% 

%%%%%%%%%%%%%%%%%%%%%%%%%%%%%%%%%%%%%%%%%%%%%%%%%%%%%%%%%%%%%%%%%%%%%%
\subsection{Bell Inequality Violation Results}
\label{sec:results_bell}

Table~\ref{tab:CHSH_results} (top) presents results for Bell inequality violation at parton-level.  Bell inequality violation is measured by $(B-\sqrt{2})$ where $B$ is given by Eq.~\eqref{eq:CHSHinHelicity}.
Results with the individual method are calculated from Eq.~\eqref{eq:cij_individual} and with the direct method from Eq.~\eqref{eq:b_direct}.  Bell inequality violation occurs when $(B-\sqrt{2})>0$.
With only statistical uncertainties, we find that Bell inequality violation can only be probed at $\approx 2\sigma$ with $300~{\rm fb}^{-1}$.  With the projected luminosity of the HL-LHC the significance is above $5\sigma$.

Bell inequality violation at detector-level after parametric fitting is shown in Table~\ref{tab:CHSH_results} (bottom).  The individual measurements have an uncertainty that increases by a factor of $1.6$ compared to the parton-level results.  The direct measurements, which use Eq.~\eqref{eq:CHSHbyFit}, on the other hand increase by a factor of $2.3$ and are actually worse than the individual measurements.  This is because the uncertainty depends on the shape of the distribution and the properties of the detector smearing.   With $300~{\rm fb}^{-1}$ the significance is only $1.3\sigma$ and even at the HL-LHC the significance only reaches $4.1\sigma$.

\begin{figure} [tb]
  \begin{center}
  \includegraphics[width=0.47\textwidth]{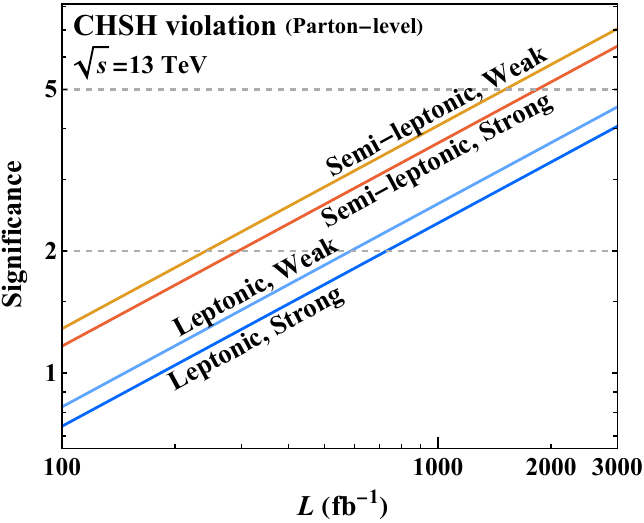}
\quad\quad
  \includegraphics[width=0.47\textwidth]{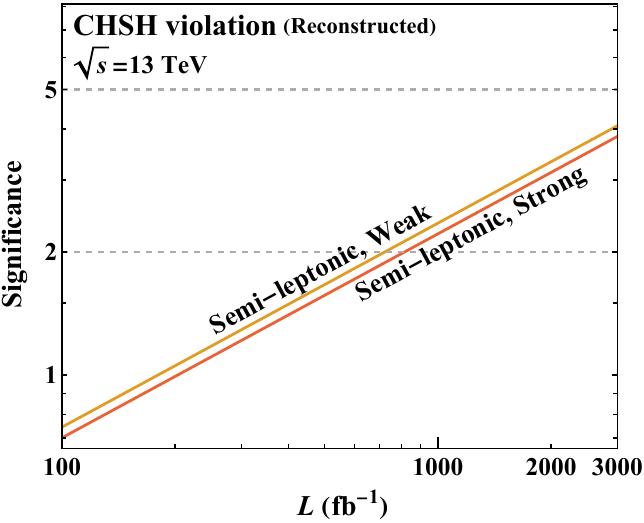}
    \caption{Expected significance of CHSH violation detection as a function of the integrated luminosity at the $13~{\rm TeV}$ LHC at parton-level (left) and after detector simulation, reconstruction, and parametric fitting (right).}
  \label{fig:CHSH_results}
  \end{center}
\end{figure}

The significance as a function of luminosity is shown in Fig.~\ref{fig:CHSH_results} (left) at parton-level.  We show results from the leptonic channel for comparison.  With the estimation from Sec.~\ref{sec:sketch}, we expected a $60\%$ improvement over the leptonic result at the parton level, while we obtain a $54\%$ improvement. Our leptonic result is consistent with Ref.~\cite{Aguilar-Saavedra:2022uye} (see Appendix~\ref{app:comparison} for a full comparison).  The detector-level result is shown in Fig.~\ref{fig:CHSH_results} (right).  Comparing to the detector-level leptonic~\cite{Severi:2021cnj} results we find a factor of $3$ improvement thanks to the higher efficiency in our channel (see Appendix~\ref{app:comparison}). 

\begin{table} [tb]
\centering
\begin{tabular}{||c||c c c c||c c||} 
 \hline \xrowht{3.8mm}
\hspace{0.15em} \multirow{2}{5.3em}{Parton-level} \hspace{0.15em} & \multirow{2}{4em}{Efficiency} & $\epsilon N_{\mathrm{parton}}$ & \multicolumn{2}{c||}{$B-\sqrt{2}$} & \multicolumn{2}{c||}{Significance} \\
  & & $(300\, \mathrm{fb}^{-1})$ & (Individual) & (Direct) & $(300\, \mathrm{fb}^{-1})$ & $(3000\, \mathrm{fb}^{-1})$ \\
 \hline
 Weak & 0.080 & 6280 & $0.22 \pm 0.11$ & $0.22 \pm 0.10$ & $2.2\sigma$ & $7.0\sigma$  \\
 \hline
 Strong & 0.078 & 4127 & $0.26 \pm 0.14$ & $0.25 \pm 0.12$ & $2.0\sigma$ & $6.4\sigma$ \\ 
 \hline
\end{tabular}
\\[1mm]
\begin{tabular}{||c||c c c||c c||} 
 \hline \xrowht{3.8mm}
 \multirow{2}{6.3em}{Reconstructed} & \hspace{2.85em}$N_{\mathrm{detected}}$\hspace{2.85em}  & \multicolumn{2}{c||}{$B-\sqrt{2}$} & \multicolumn{2}{c||}{Significance} \\
  & $(300\, \mathrm{fb}^{-1})$ &  (Individual)  & (Direct) & $(300\, \mathrm{fb}^{-1})$ & $(3000\, \mathrm{fb}^{-1})$ \\
 \hline
 Weak & 6280 & $0.23 \pm 0.18$ & $0.22 \pm 0.22$ & $1.3\sigma$ & $4.1\sigma$  \\
 \hline
 Strong & 4127 & $0.27 \pm 0.22$ & $0.25 \pm 0.28$ & $1.2\sigma$ & $3.8\sigma$ \\ 
 \hline
\end{tabular}
\caption{Measurements of $(B-\sqrt{2})$ generated at $\sqrt{s} = 13~{\rm TeV}$ and $\mathcal{L}=139~{\rm fb}^{-1}$ at parton-level (top) and after detector simulation, reconstruction, and parametric fitting (bottom).  CHSH violation is indicated by $(B-\sqrt{2})>0$.  The efficiency indicated is the average over the specified signal region.  The significance uses the direct measurement at parton-level and uses the individual measurement at reconstruction-level.} 
\label{tab:CHSH_results}
\end{table}

%%%%%%%%%%%%%%%%%%
%%%%%%%%%%%%%%%

%%%%%%%%%%%%%%%%%%%%%%%%%%%%%%%%%%%%%%%%%%%%%%%%%%%%%%%%%%%%%%%%%%%%%%
%%%%%%%%%%%%%%%%%%%%%%%%%%%%%%%%%%%%%%%%%%%%%%%%%%%%%%%%%%%%%%%%%%%%%%
\section{Summary and Conclusions}
\label{sec:conclusion}

There has been increasing interest in testing quantum entanglement and violations of Bell inequalities at high-energy colliders, which explore physics at much shorter space-time scales than traditional quantum experiments.  The $t\bar t$ system is an exemplar of a two qubit system where the detailed quantum mechanical properties of the system are exhibited through the production and decay of the $t$ and $\bar{t}$.  In this article, we explored entanglement in the $t\bar{t}$ system at the LHC via spin correlations when one of the top quarks decays leptonically and the other hadronically.  This channel has advantages over the fully leptonic channel, namely that there are roughly six times more events and the kinematic reconstruction is more efficient.

In Sec.~\ref{sec:qm}, after a brief review of quantum entanglement and Bell non-locality, we identified observables to test these quantum properties.  These quantum observables were related to collider observables in Sec.~\ref{sec:ttbar}.  In particular the spins of the $t$ and $\bar{t}$ are the qubits while spin correlations encode the entanglement between qubits.
The spins are then measured through the angles of the decay products of the $t$ and $\bar{t}$.

In Sec.~\ref{sec:lhc}, we showed our results in searching for evidence of quantum entanglement and Bell inequality violation in the semi-leptonic decay channel where the final state includes one lepton, one neutrino, two $b$-jets, and two light-quark-initiated jets from the $W$ decay.  The $t\bar{t}$ system exhibits entanglement both near threshold and at high $p_T$.  We showed that the events near threshold provide a more sensitive probe of quantum entanglement owing to a larger number of events relative to the high-$p_T$ region.  Tests of Bell inequality violation, on the other hand, require a stronger signal which is only present in the signal region with highly-boosted top quarks.

The semi-leptonic channel, which is the focus of this work, yields a higher efficiency for event reconstruction than the leptonic case.  Going beyond just the parton-level analysis, we performed a detector simulation, followed by parametric fitting to correct the detailed angular observables.  We found that this approach leads to a more stable outcome than the practice of unfolding.  As a result, the sensitivity for quantum entanglement detection is expected to be 60\% better than in the leptonic channel.  In $139~{\rm fb}^{-1}$ ($3~{\rm ab}^{-1}$) of data at the LHC (HL-LHC), it should be feasible to measure entanglement at a precision of $\lesssim 3\%\ (0.7\%)$ which is shown in Table~\ref{tab:concurrence_results} and in Fig.~\ref{fig:concurrence_results}.

The same expectation of $60\%$ improvement applies to Bell inequality violation detection. When compared to previous leptonic studies, the improvement reached a factor of $3$ better than for the leptonic channel due to a substantially higher reconstruction efficiency we achieved.  The overall detection of Bell inequality violation, however, is still challenging.  With $300~{\rm fb}^{-1}$ ($3~{\rm ab}^{-1}$) integrated luminosity at the LHC Run-3 (HL-LHC), we expect a sensitivity of $1.3 \sigma$ ($4.1 \sigma$) as shown in Table~\ref{tab:CHSH_results} and Fig.~\ref{fig:CHSH_results}.  A full comparison between previous results is shown in Appendix~\ref{app:comparison}.

In summary, we demonstrated that the semi-leptonic decay of the $t\bar{t}$ system is the premier channel for testing entanglement and Bell inequality violation at the LHC.  Performing a detector simulation and correcting the results with parametric fitting are indispensable components of an accurate prediction.  We project that at the HL-LHC entanglement can be measured nearly to the percent-level and that strong evidence will be obtained for Bell inequality violation.  There are a number of future directions such as studying the fully hadronic decay channel of the $t\bar{t}$ system and describing the small backgrounds in a quantum mechanical framework.  The LHC is a promising environment to study quantum mechanics at the TeV scale.

%%%%%%%%%%%%%%%%%%%%%%%%%%%%%%%%%%%%%%%%%%%%%%%%%%%%%%%%%%%%%%%%%%%%%%
%%%%%%%%%%%%%%%%%%%%%%%%%%%%%%%%%%%%%%%%%%%%%%%%%%%%%%%%%%%%%%%%%%%%%%
\section*{Acknowledgements}

The authors would like to thank Mikael Kuusela for detailed discussion on unfolding, Joseph Boudreau and Kun Cheng for useful discussions, and Ze Chen for computing assistance.  This work was supported in part by the U.S.~Department of Energy under grant Nos.~DE-SC0007914 and in part by the Pitt PACC.  TH would like to thank the Aspen Center for Physics, where part of this work is complete, which is supported by the National Science Foundation (NSF) grant PHY-1607611.  ML is also supported by the National Science Foundation under grant no. PHY-2112829

%%%%%%%%%%%%%%%%%%%%%%%%%%%%%%%%%%%%%%%%%%%%%%%%%%%%%%%%%%%%%%%%%%%%%%
%%%%%%%%%%%%%%%%%%%%%%%%%%%%%%%%%%%%%%%%%%%%%%%%%%%%%%%%%%%%%%%%%%%%%%
\appendix

%%%%%%%%%%%%%%%%%%%%%%%%%%%%%%%%%%%%%%%%%%%%%%%%%%%%%%%%%%%%%%%%%%%%%%
%%%%%%%%%%%%%%%%%%%%%%%%%%%%%%%%%%%%%%%%%%%%%%%%%%%%%%%%%%%%%%%%%%%%%%
\section{Unfolding and Parametric Fitting}
\label{app:unfolding}

Consider a distribution $\vec{x}_{\rm truth}$ that is produced at a collider experiment.  For example, the invariant mass spectrum or energy spectrum of a particle.  This underlying distribution is not measured directly because the detector itself has limitations and resolutions which result in smearing.  Thus the detected distribution is $\vec{x}_{\rm detected}$.

The truth and detected distributions can be related by the forward process which can be called ``folding''~\cite{Blobel:1984ku,Blobel:2011fih,Schmitt:2016orm}:
\begin{equation} \label{eq:folding}
\vec{x}_{\rm truth}
\xrightarrow{\text{folding}} 
\vec{x}_{\rm detected} = R \cdot \vec{x}_{\rm truth},
\end{equation}
where the matrix $R$ is the response matrix that describes the effects of detector smearing and phase space cuts.

Only $\vec{x}_{\rm detected}$ is measured, but we require $\vec{x}_{\rm truth}$ to extract the underlying physics parameters.  To do this, first,  $\vec{x}_{\rm truth}$ is generated by Monte Carlo.  Then a detector simulation can produce $\vec{x}_{\rm detected}$ from $\vec{x}_{\rm truth}$ which allows us to compute $R$ from Monte Carlo.  Given $R$ we can make an estimate of $\vec{x}_{\rm truth}$ that corresponds to some detected data.

%%%%%%%%%%%%%%%%%%%%%%%%%%%%%%%%%%%%%%%%%%%%%%%%%%%%%%%%%%%%%%%%%%%%%%
\subsection{Unfolding}

Unfolding is the mathematical procedure of inverting Eq.~\eqref{eq:folding} in order to solve for $\vec{x}_{\rm truth}$:
\begin{equation} \label{eq:unfolding}
\vec{x}_{\rm unfolded} = R^{-1} \cdot \vec{x}_{\rm detected}.
\end{equation}
Once $\vec{x}_{\rm unfolded}$ is obtained, the underlying physics parameters $\Theta$ can be extracted through a fit, asymmetry measurement, etc.

The response matrix $R$ quantifies the detector smearing and the loss of events which do not pass phase space cuts, and is therefore often an ill-conditioned matrix.  To find a stable inversion of $R$, one typically needs to apply regularization where ambiguity arises when choosing the form and the strength of the regularization.   That is why we write $\vec{x}_{\rm unfolded}$ in Eq.~\eqref{eq:unfolding} rather than $\vec{x}_{\rm truth}$.  As explained in Ref.~\cite{kuusela2012statistical}, without a careful choice of regularization strength one may induce a bias and underestimate the uncertainty.  Recent methods have been proposed to avoid such subtleties~\cite{Stanley_2022}. 

The bias quantifies how far the unfolded distribution $\vec{x}_{\rm unfolded}$ is from a true inversion of the response matrix applied to $x_{\rm detected}$.  When $R$ is ill-conditioned, some bias is necessary but a large bias indicates that $\vec{x}_{\rm unfolded}$ does not accurately describe $\vec{x}_{\rm truth}$.  The variance measures how much the unfolded distribution changes with respect to statistically different detected data.

%%%%%%%%%%%%%%%%%%
\begin{table} [tb]
\begin{center}
\begin{tabular}{|ccc|}
\hline
Method                       & Package                 & Regularization   \\
& &Parameters
\\ \hline
Iterative Bayesian (IB)          & {\tt RooUnfoldBayes} \cite{Adye:2011gm}  & $n_I$  \\
Singular Value Decomposition (SVD) & {\tt RooUnfoldSvd} \cite{Adye:2011gm}    & $\tau$ or $m$  \\
One-at-a-time Strict Bounds (OSB)                     & Ref.~\cite{Stanley_2022}            & $-$  \\
\hline
\end{tabular}
\caption{Unfolding algorithms and their regularization parameters.}
\label{tab:unfolding_methods}
\end{center}
\end{table}
%%%%%%%%%%%%%%%%%%

We list several unfolding algorithms in Table~\ref{tab:unfolding_methods} along with the package we use for their implementation and their regularization parameters.  The Iterative Bayesian (IB) method~\cite{dagostini2010improved} is regularized by the number of iterations $n_I$.  The Singular Value Decomposition (SVD) method~\cite{Hocker:1995kb} is parametrized by $\tau$ which is the coefficient of the regularization term.  The value of $\tau$ is often set by the square of $m$th singular value (in descending order) of a matrix related to the second derivative of the truth distribution.  Both of these methods are commonly used in theory studies.  The One-at-a-time Strict Bound (OSB) method, on the other hand, is not commonly used, but is free from any regularization~\cite{Stanley_2022}. Instead, the inputs are general constraints on the expected shape of the unfolded distribution.

To compare methods we consider the two quantities: $\cos{\theta^\cA_n}\cos{\theta^\cB_n}$ and $\cos{\theta^\cA_r}\cos{\theta^\cB_r}$.  Events are restricted to the weak signal region described in Sec.~\ref{sec:regions} which is relevant for Bell inequality violation.  We use an integrated luminosity of $300~{\rm fb}^{-1}$.  The functional form of the truth distribution is given in Eq.~\eqref{eq:cosAcosB}.

\begin{figure} [tb]
  \begin{center}
  \includegraphics[width=0.45\textwidth] {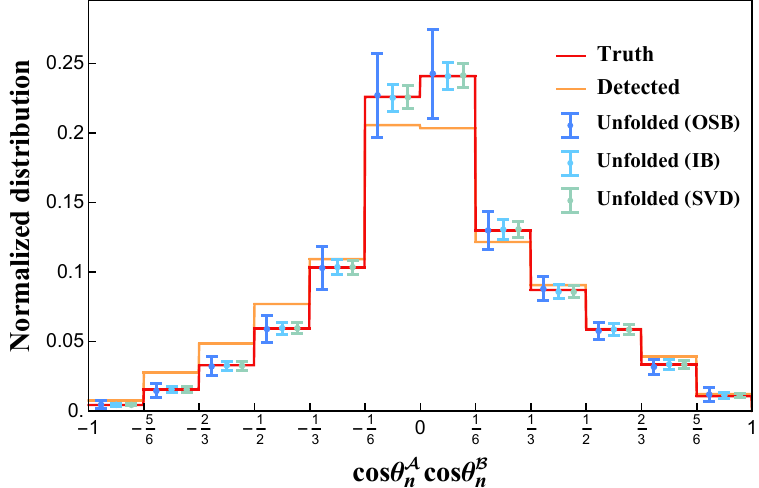}
  \quad\quad\quad
  \includegraphics[width=0.45\textwidth] {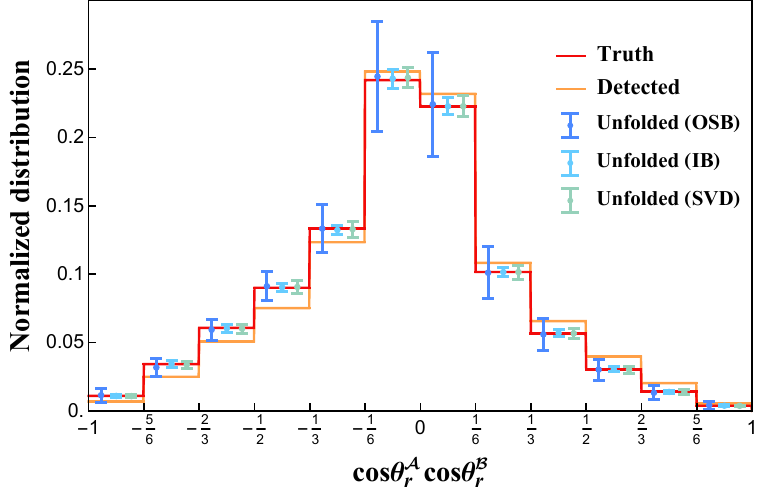}
  \caption{Distributions of $\cos{\theta^\cA_n}\cos{\theta^\cB_n}$ (left) and $\cos{\theta^\cA_r}\cos{\theta^\cB_r}$ (right) for parton-level truth data (red), for detector-level data (orange), and after applying the unfolding methods OSB (blue), IB (light blue), and SVD (green), computed at $\sqrt{s}=13~{\rm TeV}$.}
  \label{fig:distribution_cosAncosBn_cosArcosBr}
  \end{center}
\end{figure}

\begin{figure} [tb]
  \begin{center}
  \includegraphics[width=0.42\textwidth] {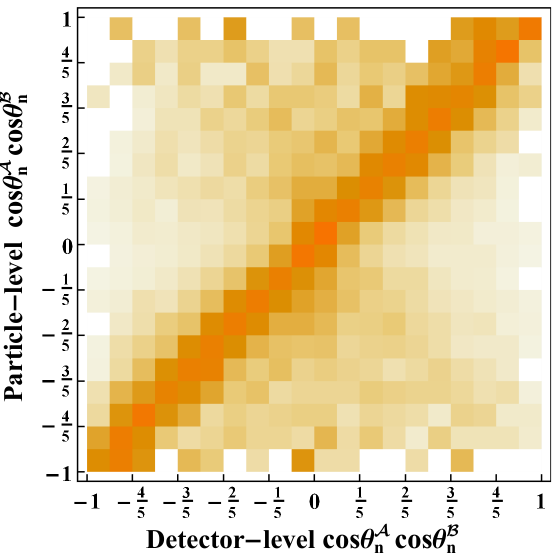}
  \quad\quad\quad
  \includegraphics[width=0.49\textwidth] {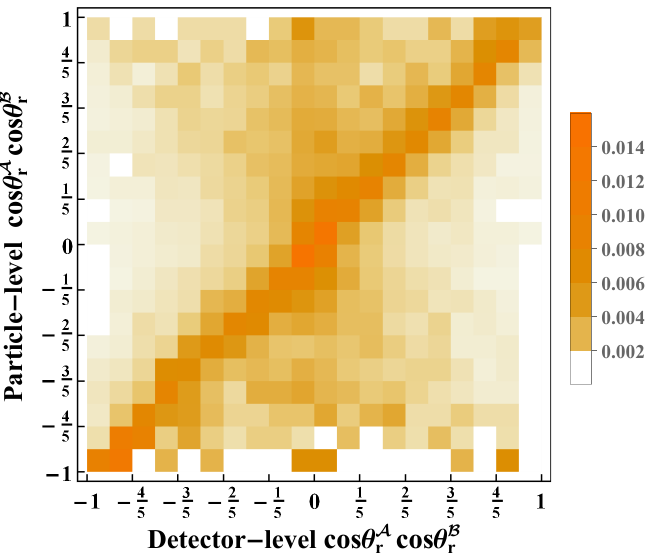}
  \caption{Response matrix of $\cos{\theta^\cA_n}\cos{\theta^\cB_n}$ (left) and $\cos{\theta^\cA_r}\cos{\theta^\cB_r}$ (right) computed at $\sqrt{s}=13~{\rm TeV}$.  Parton-level events have signal region cuts and no event selection, while detector-level events include detector effects and the effects of event selection described in Sec.~\ref{sec:simulation}.}
  \label{fig:response_matrices}
  \end{center}
\end{figure}

In Fig.~\ref{fig:distribution_cosAncosBn_cosArcosBr} we show the distribution at parton-level with only signal region cuts (red) and after detector effects and event selection cuts (orange).  The uncertainties are determined by calculating the variance from performing the same calculation in different instances of the same dataset, {\it i.e.} running pseudo-experiments.
The response matrices for these processes are shown in Fig.~\ref{fig:response_matrices}.

%%%%%%%%%%%%%%%%%%
%%%%%%%%%%%%%%%

Figure~\ref{fig:distribution_cosAncosBn_cosArcosBr} also shows the unfolding methods: OSB (blue), IB (light blue), and SVD (green).  For the OSB method over the full domain we require the unfolded distribution to be positive and separately over negative and positive input values we require the unfolded distribution to be monotonic and convex.  We follow the aggregation strategy of starting with an initial value $n_{\rm bin} = 48$, before aggregating these into larger bins.  For the IB method we use $n_I=4$ and $n_{\rm bins}=12$ while for the SVD method we use $m=4$ and $n_{\rm bins}=12$.  The results show a very stable central value for all the methods, however, the uncertainty varies substantially between methods.  The OSB method does not have free parameters while the IB and SVD methods do have free parameters.  We investigate the dependence on these parameters further below.

%%%%%%%%%%%%%%%%%%
%%%%%%%%%%%%%%%

%%%%%%%%%%%%%%%%%%
\begin{table} [tb]
\centering
\begin{tabular}{|c|c|c|c|c|c| } \hline
 parameter & $n_{\rm bin}$ & IB $n_I=3$ & IB $n_I=4$ & IB $n_I=5$ & IB $n_I=6$  \\
\hline
\multirow{2}{2em}{$C_{nn}$} & 6 & $0.752\pm 0.086$ & $0.752\pm 0.110$ & $0.752\pm 0.131$ & $0.753\pm 0.151$ \\ 
& 12 & $0.749\pm 0.088$ & $0.748\pm 0.111$ & $0.748\pm 0.131$ & $0.748\pm 0.150$ \\ 
\hline

\multirow{2}{2em}{$C_{rr}$} & 6 & $-0.895\pm 0.064$ & $-0.894\pm 0.083$ & $-0.894\pm 0.102$ & $-0.892\pm 0.117$ \\ 
& 12 & $-0.893\pm 0.062$ & $-0.892\pm 0.081$ & $-0.891\pm 0.099$ & $-0.890\pm 0.115$ \\ 
\hline

\multirow{2}{7em}{$C_{nn}-C_{rr}-\sqrt{2}$} & 6 & $0.232\pm 0.107$ & $0.232\pm 0.138$ & $0.232\pm 0.166$ & $0.230\pm 0.191$ \\ 
& 12 & $0.227\pm 0.108$ & $0.226\pm 0.137$ & $0.224\pm 0.165$ & $0.223\pm 0.189$ \\ 
\hline

\end{tabular}
\caption{Parameter estimation via unfolding with the IB method.} 
\label{tab:unfolding_IB}
\end{table}
%%%%%%%%%%%%%%%%%%

Intuitively, as the regularization strength increases, more bias is introduced but the variance decreases.  When the regularization strength decreases, the bias is reduced but the variance increases.  In Table~\ref{tab:unfolding_IB} we vary the regularization parameter $n_I$ and the number of bins $n_{\rm bin}$ using the IB method.  We show results for measuring $C_{nn}$, $C_{rr}$, and the combination $C_{nn}-C_{rr}-\sqrt{2}$.  We find that the unfolded central values are stable under variations in both $n_I$ and $n_{\rm bin}$.  The uncertainty, on the other hand, is stable under changes in $n_{\rm bin}$, but varies by up to $75\%$ while changing $n_I$.  Larger $n_I$ reduces the regularization which is why the uncertainty increases with $n_I$.

%%%%%%%%%%%%%%%%%%
\begin{table} [tb]
\centering
\begin{tabular}{|c|c|c|c|c|c| } \hline
 parameter & $n_{\rm bin} $ & SVD $m=3$ & SVD $m=4$ & SVD $m=5$ & SVD $m=6$  \\
\hline
\multirow{2}{2em}{$C_{nn}$} & 6 & $0.749\pm 0.132$ & $0.749\pm 0.161$ & $0.749\pm 0.184$ & $0.750\pm 0.199$ \\ 
& 12 & $0.746\pm 0.115$ & $0.748\pm 0.136$ & $0.748\pm 0.152$ & $0.749\pm 0.169$ \\ 
\hline

\multirow{2}{2em}{$C_{rr}$} & 6 & $-0.892\pm 0.165$ & $-0.892\pm 0.230$ & $-0.900\pm 0.260$ & $-0.897\pm 0.303$ \\ 
& 12 & $-0.894\pm 0.142$ & $-0.899\pm 0.189$ & $-0.900\pm 0.209$ & $-0.899\pm 0.245$ \\ 
\hline

\multirow{2}{7em}{$C_{nn}-C_{rr}-\sqrt{2}$} & 6 & $0.226\pm 0.211$ & $0.227\pm 0.280$ & $0.235\pm 0.318$ & $0.232\pm 0.363$ \\ 
& 12 & $0.226\pm 0.167$ & $0.232\pm 0.219$ & $0.233\pm 0.247$ & $0.234\pm 0.287$ \\ 
\hline

\end{tabular}
\caption{Parameter estimation via unfolding with the SVD method.} 
\label{tab:unfolding_SVD}
\end{table}
%%%%%%%%%%%%%%%%%%

Table~\ref{tab:unfolding_SVD} shows results using the SVD method while varying the regularization parameter $m$ and the number of bins $n_{\rm bin}$.  Again, the central value is stable with respect to changes in $m$ and $n_{\rm bin}$, but the uncertainty changes with $n_{\rm bin}$ and with $m$ up to 75\%.  A larger value of $m$ means taking a smaller squared singular value which corresponds to less regularization.

%%%%%%%%%%%%%%%%%%%%%%%%%%%%%%%%%%%%%%%%%%%%%%%%%%%%%%%%%%%%%%%%%%%%%%
\subsection{Parametric Fitting}

When only the extracted physics parameter $\Theta$ is required and not the full distribution $\vec{x}_{\rm truth}$ one can calculate the dependence of the truth distribution on the parameter $\Theta$.  This is the method more commonly used by experimentalists~\cite{Blobel:1984ku,Blobel:2011fih,Schmitt:2016orm} and in this work we will call it ``parametric fitting.''  This is sometimes called template fitting when the functional dependence of $\Theta$ is unknown and template distributions are used.

Writing the truth distribution as a function of $\Theta$ we have
\begin{equation}
\vec{x}_{\rm truth}(\Theta)
\xrightarrow{\text{folding}} 
\vec{x}_{\rm predicted}(\Theta) = R \cdot \vec{x}_{\rm truth}(\Theta).
\end{equation}
The parameter $\Theta$ is now extracted by fitting $\vec{x}_{\rm predicted}(\Theta)$ to $\vec{x}_{\rm detected}$.

We perform this parameter extraction by a binned maximum likelihood fit where the likelihood function is
\begin{align}
L(\Theta) &= \prod_{\alpha=1}^{\rm n_{\rm bins}}
{\rm Poisson}\left( x_{{\rm detected},\alpha}, x_{{\rm predicted},\alpha}(\Theta)\right), \\
&= \prod_{\alpha=1}^{\rm n_{\rm bins}}
{\rm Poisson}\left( x_{{\rm detected},\alpha}, \sum_\beta R_{\alpha\beta} x_{{\rm truth},\beta}(\Theta)\right), 
\end{align}
where ${\rm Poisson}(x,\lambda)$ is the Poisson distribution for random variable $x$ with mean $\lambda$.
The response matrix $R$ is calculated from simulation, the distribution $x_{\rm truth}(\Theta)$ as a function of $\Theta$ is known analytically in all cases that we study.  For example, for $\Theta=C_{ij}$, the truth distribution is given by Eq.~\eqref{eq:cosAcosB}.  To obtain $\Theta$ we maximize the logarithm of the likelihood function.

As with unfolding, the uncertainty is calculated by performing pseudo-experiments.  When varying the number of bins ($n_{\rm bin} = 5, 10, 20$) we find the uncertainty changes by less than 5\%.

%%%%%%%%%%%%%%%%%%
\begin{table}[tb]
\centering
\begin{tabular}{|c|c|c|c| } \hline
& Truth & OSB Unfolding & Parametric Fitting \\
\hline
{$C_{nn}$} &  $0.754 \pm 0.079$ & $0.748\pm 0.370$ & $0.754 \pm 0.116$ \\
\hline

{$C_{rr}$} &  $-0.884 \pm 0.079$ & $-0.890\pm 0.472$ & $-0.892 \pm 0.137$ \\
\hline

{$C_{nn}-C_{rr}-\sqrt{2}$} & $0.224 \pm 0.112$ & $0.223 \pm 0.600$ & $0.231 \pm 0.179$ \\
\hline

\end{tabular}
\caption{Parameter estimation via OSB unfolding and parametric fitting computed at $\sqrt{s}=13~{\rm TeV}$ in the weak signal region. The uncertainties on the truth results are statistical.}
\label{tab:unfolding_OSB_parametric_fitting}
\end{table}
%%%%%%%%%%%%%%%%%%

Table~\ref{tab:unfolding_OSB_parametric_fitting} contrasts the results from parametric fitting with OSB unfolding.  The truth result is used as a baseline where there are no smearing effects, but the number of events used to determine the uncertainty is rescaled by the average reconstruction efficiency.  While OSB unfolding does not have a dependence on regularization the resulting uncertainties are substantially larger than the statistical uncertainties.  Parametric fitting also does not depend on regularization and increases the uncertainty, but by a more modest amount.  The increase in uncertainty depends on the detector smearing, the phase space cuts, and the form of the expected distribution for the parameter.  For this reason, each fitted parameter has a different increase in uncertainty relative to the statistics only uncertainty.  In Table~\ref{tab:unfolding_OSB_parametric_fitting} the increase is about a factor of $1.4 - 1.7$, while for concurrence it is a factor of $1.9 - 3.4$ and for Bell inequality violation it is a factor of $1.6 - 2.2$.

%%%%%%%%%%%%%%%%%%%%%%%%%%%%%%%%%%%%%%%%%%%%%%%%%%%%%%%%%%%%%%%%%%%%%%
%%%%%%%%%%%%%%%%%%%%%%%%%%%%%%%%%%%%%%%%%%%%%%%%%%%%%%%%%%%%%%%%%%%%%%
\section{Comparison to Previous Results}
\label{app:comparison}

As a validation step, we compare our results with the parton-level results in Ref.~\cite{Aguilar-Saavedra:2022uye} and the detector-level results in Ref.~\cite{Severi:2021cnj}.  Our results are for the semi-leptonic channel and we use the event selection specified in Sec.~\ref{sec:simulation}.  For the purposes of comparison we do not use our signal regions but instead use the signal regions from Ref.~\cite{Aguilar-Saavedra:2022uye} and Ref.~\cite{Severi:2021cnj}.

The parton-level comparison is shown in Table~\ref{tab:comparison_parton_level}.  We apply an efficiency of 0.12 and use a luminosity of $139~{\rm fb}^{-1}$ to match Ref.~\cite{Aguilar-Saavedra:2022uye}.  The central values agree relatively well.  The small differences may result from using different PDF sets~\cite{Aguilar-Saavedra:2022uye}.  As estimated in Sec.~\ref{sec:sketch} our uncertainties should be about $60\%$ smaller than the leptonic results.  The table confirms this is an accurate estimation.

%%%%%%%%%%%%%%%%%%
\begin{table} [tb]
\begin{center}
\begin{tabular}{|c|c|c|c|c|}
\hline
    Observable & \multicolumn{3}{c|}{Entanglement: $|C_{rr}+C_{kk}|-C_{nn}-1$} & CHSH: $(C_{rr}-C_{nn})-\sqrt{2}$ \\
\hline
    Region & Threshold $\slashed{\beta}$ & Threshold $\beta$ & Boosted  & Boosted \\
\hline
   Ref.~\cite{Aguilar-Saavedra:2022uye}  &  $0.560\pm 0.020$ & $0.680\pm 0.022$ & $0.671\pm 0.069$ & $0.218\pm 0.141$ \\
\hline
   This work  &  $0.529\pm 0.013$ & $0.634\pm 0.015$ & $0.650\pm 0.042$ & $0.212\pm 0.085$ \\
\hline
\end{tabular}
\caption{Parton-level comparison between leptonic results from Ref.~\cite{Aguilar-Saavedra:2022uye} with semi-leptonic results from this work.  The semi-leptonic channel is expected to have uncertainties that are $60\%$ smaller.}
\label{tab:comparison_parton_level}
\end{center}
\end{table}
%%%%%%%%%%%%%%%%%%

The detector-level comparison is shown in Table~\ref{tab:comparison_detector_level}.  We use a luminosity of $139~{\rm fb}^{-1}$ for entanglement and $350~{\rm fb}^{-1}$ for CHSH violation to match Ref.~\cite{Severi:2021cnj}.  In the threshold region the efficiencies are similar: their leptonic sample has an efficiency of 0.08 while our semi-leptonic sample has an efficiency of 0.12.  In the high-$p_T$ region their leptonic sample has an efficiency of 0.011 (taken from Appendix B of Ref.~\cite{Aguilar-Saavedra:2022uye}) while our semi-leptonic sample has an efficiency of 0.08.  The higher efficiency in the semi-leptonic sample is expected.

For entanglement, the central values are similar, but not quite matching.  Our central values in these regions, however, do match with those from Ref.~\cite{Aguilar-Saavedra:2022uye}.  For the ``threshold, strong'' region our uncertainty is larger by a factor of $2$.  In this region the unfolding adds no uncertainty in Ref.~\cite{Severi:2021cnj} while in our work the parametric fitting always increases the uncertainty by a factor of $1.6 - 3$.  Accounting for the $60\%$ improvement from statistics in the semi-leptonic these results are consistent.  In the ``high-$p_T$, strong'' region our uncertainty is lower by $25\%$.  The unfolding from Ref.~\cite{Severi:2021cnj} increased the uncertainty by about a factor of $1.5$.  For CHSH violation, the central values are consistent.  In the ``high-$p_T$, strong'' our uncertainty is a factor of $3.4$ smaller.  While the unfolding from Ref.~\cite{Severi:2021cnj} still only increases the statistical uncertainty by a factor of $1.5$, the reconstruction efficiency in our sample is much higher.

Finally, we briefly compare to Ref.~\cite{Dong:2023xiw}.  They provide detector-level results which include a deep neural network reconstruction algorithm and SVD unfolding. Our weak signal region from Sec.~\ref{sec:regions} has approximately a factor of $3$ times more events than the signal region used in Ref.~\cite{Dong:2023xiw}.  In addition, our parametric fitting increases the statistical uncertainty by a factor of roughly $3$ while in Ref.~\cite{Dong:2023xiw} the unfolding decreases the statistical uncertainty slightly. 

%%%%%%%%%%%%%%%%%%
\begin{table} [tb]
\begin{center}
\begin{tabular}{|c|c|c|c|}
\hline \xrowht{5.5mm}
    Observable & \multicolumn{2}{c|}{Entanglement: $|C_{rr}+C_{kk}|-C_{nn}-1$} & CHSH: $(C_{rr}-C_{nn})-\sqrt{2}$ \\
\hline
    Region & Threshold, strong &  High-$p_T$, strong & High-$p_T$, strong \\
\hline
    Ref.~\cite{Severi:2021cnj}
      & $0.38\pm 0.02$ & $0.42\pm 0.10$ & $0.21\pm 0.54$ \\
\hline
    This work
      & $0.45 \pm 0.04$ & $0.58 \pm 0.08$ & $0.19 \pm 0.16$ \\
\hline
\end{tabular}
\caption{Detector-level comparison between leptonic results from Ref.~\cite{Severi:2021cnj} with the semi-leptonic results from this work.  The semi-leptonic channel is expected to have uncertainties that are 60\% smaller without accounting for differences in reconstruction efficiency.  The CHSH result from Ref.~\cite{Severi:2021cnj} is multiplied by $1/\sqrt{2}$ to match our normalization.}
\label{tab:comparison_detector_level}
\end{center}
\end{table}
%%%%%%%%%%%%%%%%%%

%%%%%%%%%%%%%%%%%%%%%%%%%%%%%%%%%%%%%%%%%%%%%%%%%%%%%%%%%%%%%%%%%%%%%%
%%%%%%%%%%%%%%%%%%%%%%%%%%%%%%%%%%%%%%%%%%%%%%%%%%%%%%%%%%%%%%%%%%%%%%
\section{Spin Analyzing Power for Hadronic Top Decays}
\label{app:spin}

Consider an ensemble of polarized top quarks with polarization vector $\vec{B}$, where $0 \leq |\vec{B}| \leq 1$.  The differential decay width of the top quark is
\begin{equation} \label{eq:diffDecay}
\frac{1}{\Gamma}\frac{d\Gamma}{d\cos\theta_v}
=
\frac{1}{2} \left( 1 + |\vec{B}| \kappa_v \cos\theta_v \right),
\end{equation}
where $\cos\theta_v = (\vec{B} \cdot \vec{v}) / (|\vec{B}| |\vec{v}|)$ for a direction $\vec{v}$ associated with the decay products.  The coefficient $\kappa_v$ is the spin analyzing power associated with the direction $\vec{v}$.

In the leptonic decay of a top quark, if $\vec{v} = \vec{p}_{\ell^+}$ then $\kappa_{\ell^+} = 1.0$.  The spin analyzing power ranges from $-1$ to $+1$, so the anti-lepton carries the maximum amount of information about the polarization of the top quark.

The spin analyzing powers of the other decay products can be calculated and, at leading order, are~\cite{Brandenburg:2002xr}
\begin{equation}
\kappa_{W^+} = 0.40,\quad\quad
\kappa_{b} = -0.40,\quad\quad
\kappa_{\nu} = -0.34.\quad\quad
\end{equation}
For the decay of anti-top quark, the spin analyzing power is equal in magnitude and opposite in sign for the corresponding anti-particles in the decay product.  In hadronic decays of the top quark, the vertex structure is the same with the replacement of $\ell^+ \to$ down-type anti-quark and $\nu \to$ up-type quark.  The complication in this case is that the down-type anti-quark cannot be distinguished from the up-type quark on an event-by-event basis.  They are both detected as jets.

Early on, the softer jet (the jet with the lower energy in the top rest frame) was used and has a spin analyzing power of $\kappa_{\rm soft}=0.50$~\cite{Brandenburg:2002xr}.  The intuition is that the down-type anti-quark tends to be emitted closer to the $b$-quark which makes it more often become the softer jet.

In Ref.~\cite{Tweedie:2014yda} it was shown that the optimal spin analyzing power uses a weighted sum of both the quark and anti-quark.   The optimal hadronic direction $\vec{p}_{\rm opt}$ is
\begin{equation} \label{eq:opt}
\vec{p}_{\mathrm{opt}} (\cos\theta_W)
= P_{d \to p_{\rm soft}}(\cos\theta_W) \, \hat{p}_{\mathrm{soft}} +
P_{d \to p_{\rm hard}}(\cos\theta_W) \, \hat{p}_{\mathrm{hard}},
\end{equation}
where $\hat{p}_{\mathrm{soft}}$ is the normalized three-momentum of the softer jet, $\hat{p}_{\mathrm{hard}}$ is the normalized three-momentum of the harder jet, and $\theta_W$ is the angle between one of the $W$ decay products and the $W$ momentum axis in the $W$ rest frame (shown in Fig.~\ref{fig:wdecay_diagram}).  
The functions $P_{d \to p_{\rm soft}}(\cos\theta_W)$ and $P_{d \to p_{\rm hard}}(\cos\theta_W)$ are
\begin{align}
P_{d \to p_{\rm soft}}(\cos\theta_W)
&= \frac{f(-|\cos\theta_W|)}
{f(|\cos\theta_W|)+f(-|\cos\theta_W|)},\\
P_{d \to p_{\rm hard}}(\cos\theta_W)
&= \frac{f(|\cos\theta_W|)}
{f(|\cos\theta_W|)+f(-|\cos\theta_W|)}.
\end{align}
The function $f(\cos\theta_W)$ is the probability distribution of $\cos\theta_W$ which depends on the polarization of the $W$ boson coming from the decay of the top~\cite{Tweedie:2014yda}.  Neglecting the $b$ mass the distribution is
\begin{equation}
    f (\cos\theta_W)= \frac{3}{4} \frac{m_t^2}{m_t^2 + 2m_W^2} (1-\cos^2\theta_W) + \frac{3}{8} \frac{2m_W^2}{m_t^2 + 2m_W^2} (1-\cos\theta_W)^2.
\end{equation}
The dependence of Eq.~\eqref{eq:opt} on $\cos\theta_W$ means that the spin analyzing power also is a function of $\cos\theta_W$.  The dependence of the spin analyzing power on $\cos\theta_W$ is nearly flat~\cite{Tweedie:2014yda}.

From theory, the predicted integrated value of the spin analyzing power is $\kappa_{\rm opt} = 0.638$.  To ensure the validity of our results for entanglement and Bell inequality violation we compute the spin analyzing power from simulation.

\begin{figure} [tb]
  \begin{center}
  \includegraphics[width=0.6\textwidth]{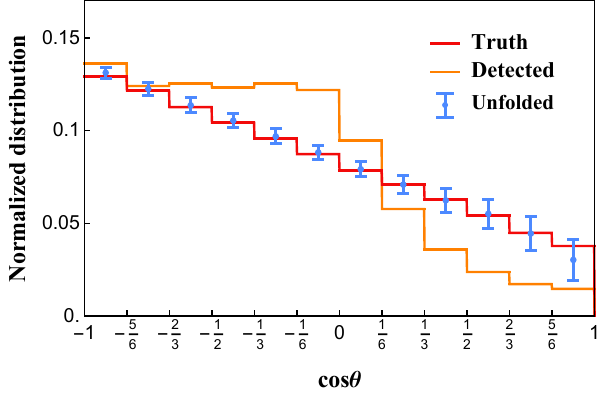}
  \caption{Differential decay width of the top quark at parton-level (red), uncorrected detector-level (orange), and after unfolding (blue markers).}
  \label{fig:distribution_costheta}
  \end{center}
\end{figure}

We use {\tt Madgraph 5}~\cite{Alwall:2011uj} to generate a sample of polarized top quarks.  The differential decay width as a function of $\cos\theta$ taken with respect to $\vec{p}_{\mathrm{opt}}$ is shown in Fig.~\ref{fig:distribution_costheta}.  The parton-level distribution is shown in red and yields a value of $\kappa_{\rm opt} = 0.640 \pm 0.004$.  The distribution at the uncorrected detector-level is shown in orange.  The blue markers indicate the distribution after unfolding and lead to a value of $\kappa_{\rm opt} = 0.654 \pm 0.037$.  Using parametric fitting yields $\kappa_{\rm opt} = 0.642 \pm 0.030$.  These numbers are summarized in Table~\ref{tab:kappa_opt}.

Having shown the robustness of the optimal hadronic spin analyzing power we use the value of $\kappa_{\rm opt} = 0.64$ in our results for entanglement and Bell inequality violation.

%%%%%%%%%%%%%%%%%%
%%%%%%%%%%%%%%%
%%%%%%%%%%%%%%%%%%
\begin{table}[H]
\centering
\begin{tabular}{|c|c|c|c|c|} \hline
& Theory & Parton-level & Unfolded & Parametric Fitted \\
\hline
$\kappa_{\rm opt}$ &  $0.638$ & $0.640\pm 0.004$ & $0.654 \pm 0.037$ & $0.642 \pm 0.030$ \\
\hline
\end{tabular}
\caption{Calculated values of the optimal hadronic spin analyzing power in top decays at theory-level, at parton-level, after unfolding, and after parametric fitting.  The uncertainties are from Monte Carlo statistics.}
\label{tab:kappa_opt}
\end{table}
%%%%%%%%%%%%%%%%%%

%%%%%%%%%%%%%%%%%%%%%%%%%%%%%%%%%%%%%%%%%%%%%%%%%%%%%%%%%%%%%%%%%%%%%%
\section{Quantum versus Fictitious States}
\label{app:qstate}

In this appendix we highlight relevant differences between quantum states and fictitious states.

%%%%%%%%%%%%%%%%%%%%%%%%%%%%%%%%%%%%%%%%%%%%%%%%
\subsection*{Non-Spin Degrees of Freedom}

The $t\bar{t}$ system at a collider is labeled by the top quark momentum in center-of-mass frame $\vec{k}$, the velocity $\vec{v}$ of the $t\bar{t}$ system relative to the lab frame, and the spins of the top and the anti-top quarks.  We denote the spin as $\ket{\alpha} = \ket{\text{spin of } t} \otimes \ket{\text{spin of }\bar{t}}$.

The spin density matrix $\rho_{\rm spin}$ for the $t\bar{t}$ system with a given $\vec{k}$ and $\vec{v}$ (which we will refer to as individual density matrix) can be written as
\begin{equation}
    \rho_{\mathrm{spin}}(\vec{k},\vec{v}) = \sum_{\alpha, \beta} \rho(\vec{k},\vec{v})_{\alpha,\beta}\ket{\alpha}\bra{\beta}.
\end{equation}
Each value of $\vec{k}$ and $\vec{v}$ yields a distinct quantum state.

%%%%%%%%%%%%%%%%%%%%%%%%%%%%%%%%%%%%%%%%%%%%%%%%
\subsection*{Quantum States}

To find the total spin density matrix for the $t\bar{t}$ system produced in a collider, we perform the sum over the phase space $\Pi$
\begin{equation}
    \rho_{\mathrm{spin}}(\Pi) = \sum_{\vec{k},\vec{v}\in \Pi} \rho_{\mathrm{spin}}(\vec{k},\vec{v}) = \sum_{\alpha, \beta} \Big(\sum_{\vec{k},\vec{v}\in \Pi} \rho(\vec{k},\vec{v})_{\alpha,\beta}\Big)\ket{\alpha}\bra{\beta}\,, \label{eq:physicalstate}
\end{equation}
where each matrix element must be evaluated in the same fixed frame. 

The $\rho_{\mathrm{spin}}(\Pi)$ obtained in this way is a ``physical'' state, in the sense that if we measure the spin observable $\mathcal{O}$, % in the lab, 
then its expectation value is simply given by $\langle \cO \rangle=\mathrm{tr}(\mathcal{O}\rho_{\mathrm{spin}}(\Pi))$.  We would also call this a genuine quantum state.

Quantum states, however, can exhibit cancellations in the entanglement of the total spin density matrix, despite entanglement among individual spin density matrices.  This cancellation occurs due to the summation over the azimuthal angle of the production plane.

For example, consider the case where $\vec{v}=0$ and each $t\bar{t}$ pair is produced with an angle $\theta$ in the $y-z$ plane with a spin correlation matrix of $C_{11} \approx -C_{22} \approx C_{33} \approx 1$ (all other entries are $0$).  The concurrence would be $\cC \approx 1$ corresponding to maximal entanglement.

By rotational symmetry around the $z$-axis, each $t\bar{t}$ pair with the same polar angle $\theta$, but different azimuthal angle $\phi$ is also maximally entangled.  The spin correlation matrix is
\begin{equation} \label{eq:spinmatrixC}
C = \begin{pmatrix}
\cos^2\!\phi\, C_{11} + \sin^2\!\phi\, C_{22} \phantom{a}& (C_{11}-C_{22})\cos\phi\sin\phi & 0\\
 (C_{11}-C_{22})\cos\phi\sin\phi \phantom{a}& \sin^2\!\phi\, C_{11} + \cos^2\!\phi\, C_{22} & 0 \\
0 & 0 & C_{33}
\end{pmatrix}.	
\end{equation}
Just as with the total spin density matrix, the total spin correlation matrix $C$ is given by the sum of $C$ in Eq.~\eqref{eq:spinmatrixC} over $\phi$.  After the summation the diagonal elements become: $(C_{11} + C_{22})/2 \approx 0$, $(C_{11} + C_{22})/2 \approx 0$, and $C_{33} \approx 1$ leading to a concurrence $\cC \approx 0$.

In this example, the sum of maximally entangled states became non-entangled, which is the case for the $t\bar{t}$ system with high $p_T$ at the LHC.  This is why in the boosted region there is no significant entanglement in the fixed beam basis.  The helicity basis, on the other hand, does exhibit large entanglement because it does not correspond to a quantum state.

%%%%%%%%%%%%%%%%%%%%%%%%%%%%%%%%%%%%%%%%%%%%%%%%
\subsection*{The Average of Concurrence}

Instead of computing the concurrence of a quantum state $\rho_{\rm spin}(\Pi)$, we can use other quantities that will not exhibit the same cancellations, and consequently will enhance the experimental detection.

This can be accomplished by computing the average of the concurrence $\overline{\cC}$ over states with different $\vec{k}$ and evaluated in the center-of-mass frame where $\vec{v}=0$:
\begin{equation}
\overline{\cC} = \sum_{\vec{k}, \vec{v}} \cC(\rho_{\rm spin}(\vec{k},0)).
\end{equation}
This should be contrasted with the concurrence of the quantum state $\cC(\sum_{\vec{k},\vec{v}}\rho_\mathrm{spin}(\vec{k},\vec{v}))$.

Since the concurrence is invariant under rotations, we can evaluate each term $\cC(\rho_{\rm spin}(\vec{k},0))$ in the helicity basis.  Using the results from Eq.~\eqref{eq:necessary1a} we find \\ $\cC(\rho_{\rm spin}(\vec{k},0)) = (1/2)\max(-C_{nn}(\vec{k}) + |C_{kk}(\vec{k}) +C_{rr}(\vec{k})  |-1,0)$ which leads to
\begin{align}
%\sum_{\vec{k},\vec{v}\in \Pi} \mathcal{C}(\rho_{\mathrm{spin},\vec{k},0})
\label{eq:average_concurrence1}
\overline{\cC}
&= \sum_{\vec{k},\vec{v}\in \Pi} \frac{1}{2}\mathrm{max}\big(-C_{nn}(\vec{k})+|C_{kk}(\vec{k})+C_{rr}(\vec{k})|-1,0\big), \\
\label{eq:average_concurrence2}
& \geq  \frac{1}{2}\mathrm{max}\Bigg(-\sum_{\vec{k},\vec{v}\in \Pi}C_{nn}(\vec{k})+\bigg|\sum_{\vec{k},\vec{v}\in \Pi}(C_{kk}(\vec{k})+C_{rr}(\vec{k}))\bigg|-1,0\Bigg), \\
\label{eq:average_concurrence3}
&=\cC(\overline{\rho}_{\mathrm{spin}}(\Pi)).
\end{align}
Going from Eq.~\eqref{eq:average_concurrence2} to Eq.~\eqref{eq:average_concurrence3} requires $\sum_{\vec{k},\vec{v}}C_{nn}<0$ and that $\sum_{\vec{k},\vec{v}}C_{rk}=\sum_{\vec{k},\vec{v}}C_{kr}$ are the only two non-vanishing off-diagonal entries of $\sum_{\vec{k},\vec{v}}C_{ij}$.  These conditions are true for both near threshold and in the boosted region.

%%%%%%%%%%%%%%%%%%%%%%%%%%%%%%%%%%%%%%%%%%%%%%%%
\subsection*{Fictitious States}

In Eq.~\eqref{eq:average_concurrence3} we define the density matrix of a ``fictitious state'' $\overline{\rho}_{\mathrm{spin}}(\Pi)$
\begin{equation} \label{eq:fictitious}
(\overline{\rho}_{\mathrm{spin}}(\Pi))_{\alpha,\beta}=\sum_{\vec{k},\vec{v}\in \Pi} \rho(\vec{k},0)_{\alpha(\vec{k}),\beta(\vec{k})},
\end{equation}
where $\alpha$ and $\beta$ denote the axes along which we measure the spin.  Each term in the summation should be evaluated in its own center-of-mass frame.  In the helicity basis, these axes depend on $\vec{k}$ which is why they are written as $\alpha(\vec{k})$ and $\beta(\vec{k})$ on the right-hand side.  On the other hand, a quantum state is given by $(\rho_{\mathrm{spin}}(\Pi))_{\alpha,\beta} = \sum_{\vec{k},\vec{v}} \rho(\vec{k},\vec{v})_{\alpha,\beta}$ (see Eq.~\eqref{eq:physicalstate}), where each term in the sum has a certain center-of-mass velocity and the spin is measured along the same axes.

There is not an obvious physical interpretation for the fictitious state in Eq.~\eqref{eq:fictitious}, however, by Eqs.~(\ref{eq:average_concurrence1} - \ref{eq:average_concurrence3}) the average concurrence $\overline{\cC}$ is greater than or equal to the concurrence of the fictitious state.  Therefore, 
\begin{equation}
\cC(\overline{\rho}_{\mathrm{spin}}(\Pi)) > 0 
\quad\quad\quad
\Rightarrow
\quad\quad\quad
\overline{\cC}>0.
\end{equation}
This means that when the concurrence of the fictitious state is positive, there exists a sub-state that is entangled.  The same argument can be applied to CHSH violation.  In the main text, concurrence and CHSH violation are measured using fictitious states.  The derivation here justifies their validity in searches for entanglement and Bell inequality violation.

%%%%%%%%%%%%%%%%%%%%%%%%%%%%%%%%%%%%%%%%%%%%%%%%%%%%%%%%%%%%%%%%%%%%%%
%%%%%%%%%%%%%%%%%%%%%%%%%%%%%%%%%%%%%%%%%%%%%%%%%%%%%%%%%%%%%%%%%%%%%%
\section{Charm Tagging}
\label{app:ctagging}

An alternative to using the optimal hadronic direction is to only consider events with charm quarks.  In this case, the down-type quark (the strange quark) can be identified as the jet that is not charm-tagged.  Let the charm-tagging efficiency be $\epsilon_c$.

Adapting Eq.~\eqref{eq:rescaling} we have
\begin{equation} 
\frac{\text{significance}\;(t\bar{t} \to \ell s)}{\text{significance}\;(t\bar{t} \to \ell\ell)}
= \frac{\kappa_s \kappa_\ell}{\kappa_\ell \kappa_\ell}
\sqrt{\frac{\epsilon_c \text{BR}(t\bar{t} \to s\ell)}{\text{BR}(t\bar{t} \to \ell\ell)}}
= 1.78\sqrt{\epsilon_c}.
\end{equation}
In order for the subset of charm-tagged semi-leptonic events to be more sensitive than all semi-leptonic events, it is necessary that $1.78 \sqrt{\epsilon_c} > 1.60$ or $\epsilon_c > 0.95$.  In several analyses, the operating point used for charm-tagging has an efficiency of $30-40\%$ with a light-quark jet mistag rate of about $5\%$~\cite{ATLAS:2018mgv,CMS:2019hve}. 

Instead of only using charm-tagged events, it may be beneficial to combine two signal regions.  The first signal region would consist of charm-tagged events and would use the strange-inferred jet, while the second signal region would consist of the rest of the semi-leptonic events and would use the optimal hadronic direction.

\bibliographystyle{jhep}
\bibliography{refs}

\providecommand{\href}[2]{#2}\begingroup\raggedright\begin{thebibliography}{10}

\bibitem{Bell:1964kc}
J.S.~Bell, \emph{{On the Einstein-Podolsky-Rosen paradox}},
  \href{https://doi.org/10.1103/PhysicsPhysiqueFizika.1.195}{\emph{Physics
  Physique Fizika} {\bfseries 1} (1964) 195}.

\bibitem{Aoude:2022imd}
R.~Aoude, E.~Madge, F.~Maltoni and L.~Mantani, \emph{{Quantum SMEFT tomography:
  Top quark pair production at the LHC}},
  \href{https://doi.org/10.1103/PhysRevD.106.055007}{\emph{Phys. Rev. D}
  {\bfseries 106} (2022) 055007}
  [\href{https://arxiv.org/abs/2203.05619}{{\ttfamily 2203.05619}}].

\bibitem{Severi:2022qjy}
C.~Severi and E.~Vryonidou, \emph{{Quantum entanglement and top spin
  correlations in SMEFT at higher orders}},
  \href{https://arxiv.org/abs/2210.09330}{{\ttfamily 2210.09330}}.

\bibitem{Afik:2020onf}
Y.~Afik and J.R.M.n.~de~Nova, \emph{{Entanglement and quantum tomography with
  top quarks at the LHC}},
  \href{https://doi.org/10.1140/epjp/s13360-021-01902-1}{\emph{Eur. Phys. J.
  Plus} {\bfseries 136} (2021) 907}
  [\href{https://arxiv.org/abs/2003.02280}{{\ttfamily 2003.02280}}].

\bibitem{Fabbrichesi:2021npl}
M.~Fabbrichesi, R.~Floreanini and G.~Panizzo, \emph{{Testing Bell Inequalities
  at the LHC with Top-Quark Pairs}},
  \href{https://doi.org/10.1103/PhysRevLett.127.161801}{\emph{Phys. Rev. Lett.}
  {\bfseries 127} (2021) 161801}
  [\href{https://arxiv.org/abs/2102.11883}{{\ttfamily 2102.11883}}].

\bibitem{Severi:2021cnj}
C.~Severi, C.D.E.~Boschi, F.~Maltoni and M.~Sioli, \emph{{Quantum tops at the
  LHC: from entanglement to Bell inequalities}},
  \href{https://doi.org/10.1140/epjc/s10052-022-10245-9}{\emph{Eur. Phys. J. C}
  {\bfseries 82} (2022) 285}
  [\href{https://arxiv.org/abs/2110.10112}{{\ttfamily 2110.10112}}].

\bibitem{Afik:2022kwm}
Y.~Afik and J.R.M.n.~de~Nova, \emph{{Quantum information with top quarks in QCD
  production}},  \href{https://arxiv.org/abs/2203.05582}{{\ttfamily
  2203.05582}}.

\bibitem{Aguilar-Saavedra:2022uye}
J.A.~Aguilar-Saavedra and J.A.~Casas, \emph{{Improved tests of entanglement and
  Bell inequalities with LHC tops}},
  \href{https://doi.org/10.1140/epjc/s10052-022-10630-4}{\emph{Eur. Phys. J. C}
  {\bfseries 82} (2022) 666}
  [\href{https://arxiv.org/abs/2205.00542}{{\ttfamily 2205.00542}}].

\bibitem{Afik:2022dgh}
Y.~Afik and J.R.M.n.~de~Nova, \emph{{Quantum discord and steering in top quarks
  at the LHC}},  \href{https://arxiv.org/abs/2209.03969}{{\ttfamily
  2209.03969}}.

\bibitem{Barger:1988jj}
V.D.~Barger, J.~Ohnemus and R.J.N.~Phillips, \emph{{Spin Correlation Effects in
  the Hadroproduction and Decay of Very Heavy Top Quark Pairs}},
  \href{https://doi.org/10.1142/S0217751X89000297}{\emph{Int. J. Mod. Phys. A}
  {\bfseries 4} (1989) 617}.

\bibitem{Mahlon:1995zn}
G.~Mahlon and S.J.~Parke, \emph{{Angular correlations in top quark pair
  production and decay at hadron colliders}},
  \href{https://doi.org/10.1103/PhysRevD.53.4886}{\emph{Phys. Rev. D}
  {\bfseries 53} (1996) 4886}
  [\href{https://arxiv.org/abs/hep-ph/9512264}{{\ttfamily hep-ph/9512264}}].

\bibitem{Stelzer:1995gc}
T.~Stelzer and S.~Willenbrock, \emph{{Spin correlation in top quark production
  at hadron colliders}},
  \href{https://doi.org/10.1016/0370-2693(96)00178-5}{\emph{Phys. Lett. B}
  {\bfseries 374} (1996) 169}
  [\href{https://arxiv.org/abs/hep-ph/9512292}{{\ttfamily hep-ph/9512292}}].

\bibitem{Parke:1996pr}
S.J.~Parke and Y.~Shadmi, \emph{{Spin correlations in top quark pair production
  at $e^{+} e^{-}$ colliders}},
  \href{https://doi.org/10.1016/0370-2693(96)00998-7}{\emph{Phys. Lett. B}
  {\bfseries 387} (1996) 199}
  [\href{https://arxiv.org/abs/hep-ph/9606419}{{\ttfamily hep-ph/9606419}}].

\bibitem{Mahlon:1997uc}
G.~Mahlon and S.J.~Parke, \emph{{Maximizing spin correlations in top quark pair
  production at the Tevatron}},
  \href{https://doi.org/10.1016/S0370-2693(97)00987-8}{\emph{Phys. Lett. B}
  {\bfseries 411} (1997) 173}
  [\href{https://arxiv.org/abs/hep-ph/9706304}{{\ttfamily hep-ph/9706304}}].

\bibitem{Mahlon:2010gw}
G.~Mahlon and S.J.~Parke, \emph{{Spin Correlation Effects in Top Quark Pair
  Production at the LHC}},
  \href{https://doi.org/10.1103/PhysRevD.81.074024}{\emph{Phys. Rev. D}
  {\bfseries 81} (2010) 074024}
  [\href{https://arxiv.org/abs/1001.3422}{{\ttfamily 1001.3422}}].

\bibitem{Dong:2023xiw}
Z.~Dong, D.~Gon\c{c}alves, K.~Kong and A.~Navarro, \emph{{When the Machine
  Chimes the Bell: Entanglement and Bell Inequalities with Boosted
  $t\bar{t}$}},  \href{https://arxiv.org/abs/2305.07075}{{\ttfamily
  2305.07075}}.

\bibitem{Barr:2022wyq}
A.J.~Barr, P.~Caban and J.~Rembieli\'nski, \emph{{Bell-type inequalities for
  systems of relativistic vector bosons}},
  \href{https://arxiv.org/abs/2204.11063}{{\ttfamily 2204.11063}}.

\bibitem{Fabbrichesi:2023cev}
M.~Fabbrichesi, R.~Floreanini, E.~Gabrielli and L.~Marzola, \emph{{Bell
  inequalities and quantum entanglement in weak gauge boson production at the
  LHC and future colliders}},
  \href{https://doi.org/10.1140/epjc/s10052-023-11935-8}{\emph{Eur. Phys. J. C}
  {\bfseries 83} (2023) 823}
  [\href{https://arxiv.org/abs/2302.00683}{{\ttfamily 2302.00683}}].

\bibitem{Aoude:2023hxv}
R.~Aoude, E.~Madge, F.~Maltoni and L.~Mantani, \emph{{Probing new physics
  through entanglement in diboson production}},
  \href{https://arxiv.org/abs/2307.09675}{{\ttfamily 2307.09675}}.

\bibitem{Bi:2023uop}
Q.~Bi, Q.-H.~Cao, K.~Cheng and H.~Zhang, \emph{{New observables for testing
  Bell inequalities in $W$ boson pair production}},
  \href{https://arxiv.org/abs/2307.14895}{{\ttfamily 2307.14895}}.

\bibitem{Barr:2021zcp}
A.J.~Barr, \emph{{Testing Bell inequalities in Higgs boson decays}},
  \href{https://doi.org/10.1016/j.physletb.2021.136866}{\emph{Phys. Lett. B}
  {\bfseries 825} (2022) 136866}
  [\href{https://arxiv.org/abs/2106.01377}{{\ttfamily 2106.01377}}].

\bibitem{Aguilar-Saavedra:2022mpg}
J.A.~Aguilar-Saavedra, \emph{{Laboratory-frame tests of quantum entanglement in
  $H \to WW$}},  \href{https://arxiv.org/abs/2209.14033}{{\ttfamily
  2209.14033}}.

\bibitem{Aguilar-Saavedra:2022wam}
J.A.~Aguilar-Saavedra, A.~Bernal, J.A.~Casas and J.M.~Moreno, \emph{{Testing
  entanglement and Bell inequalities in $H \to ZZ$}},
  \href{https://arxiv.org/abs/2209.13441}{{\ttfamily 2209.13441}}.

\bibitem{Bernal:2023ruk}
A.~Bernal, P.~Caban and J.~Rembieli\'nski, \emph{{Entanglement and Bell
  inequalities violation in $H\to ZZ$ with anomalous coupling}},
  \href{https://arxiv.org/abs/2307.13496}{{\ttfamily 2307.13496}}.

\bibitem{Ashby-Pickering:2022umy}
R.~Ashby-Pickering, A.J.~Barr and A.~Wierzchucka, \emph{{Quantum state
  tomography, entanglement detection and Bell violation prospects in weak
  decays of massive particles}},
  \href{https://doi.org/10.1007/JHEP05(2023)020}{\emph{JHEP} {\bfseries 05}
  (2023) 020} [\href{https://arxiv.org/abs/2209.13990}{{\ttfamily
  2209.13990}}].

\bibitem{Fabbrichesi:2023jep}
M.~Fabbrichesi, R.~Floreanini, E.~Gabrielli and L.~Marzola, \emph{{Stringent
  bounds on HWW and HZZ anomalous couplings with quantum tomography at the
  LHC}}, \href{https://doi.org/10.1007/JHEP09(2023)195}{\emph{JHEP} {\bfseries
  09} (2023) 195} [\href{https://arxiv.org/abs/2304.02403}{{\ttfamily
  2304.02403}}].

\bibitem{Morales:2023gow}
R.A.~Morales, \emph{{Exploring Bell inequalities and quantum entanglement in
  vector boson scattering}},
  \href{https://arxiv.org/abs/2306.17247}{{\ttfamily 2306.17247}}.

\bibitem{Aguilar-Saavedra:2023hss}
J.A.~Aguilar-Saavedra, \emph{{Post-decay quantum entanglement in top pair
  production}},  \href{https://arxiv.org/abs/2307.06991}{{\ttfamily
  2307.06991}}.

\bibitem{Altakach:2022ywa}
M.M.~Altakach, P.~Lamba, F.~Maltoni, K.~Mawatari and K.~Sakurai, \emph{{Quantum
  information and CP measurement in $H \to \tau^+ \tau^-$ at future lepton
  colliders}},  \href{https://arxiv.org/abs/2211.10513}{{\ttfamily
  2211.10513}}.

\bibitem{Ma:2023yvd}
K.~Ma and T.~Li, \emph{{Testing Bell inequality through $h\to\tau\tau$ at
  CEPC}},  \href{https://arxiv.org/abs/2309.08103}{{\ttfamily 2309.08103}}.

\bibitem{Takubo:2021sdk}
Y.~Takubo, T.~Ichikawa, S.~Higashino, Y.~Mori, K.~Nagano and I.~Tsutsui,
  \emph{{Feasibility of Bell inequality violation at the ATLAS experiment with
  flavor entanglement of B0B\textasciimacron{}0 pairs from pp collisions}},
  \href{https://doi.org/10.1103/PhysRevD.104.056004}{\emph{Phys. Rev. D}
  {\bfseries 104} (2021) 056004}
  [\href{https://arxiv.org/abs/2106.07399}{{\ttfamily 2106.07399}}].

\bibitem{Fabbrichesi:2022ovb}
M.~Fabbrichesi, R.~Floreanini and E.~Gabrielli, \emph{{Constraining new physics
  in entangled two-qubit systems: top-quark, tau-lepton and photon pairs}},
  \href{https://arxiv.org/abs/2208.11723}{{\ttfamily 2208.11723}}.

\bibitem{Privitera:1991nz}
P.~Privitera, \emph{{Decay correlations in e+ e- ---\ensuremath{>} tau+ tau- as
  a test of quantum mechanics}},
  \href{https://doi.org/10.1016/0370-2693(92)90872-2}{\emph{Phys. Lett. B}
  {\bfseries 275} (1992) 172}.

\bibitem{Dreiner:1992gt}
H.K.~Dreiner, \emph{{Bell's inequality and tau physics at LEP}},  in \emph{{2nd
  Workshop on Tau Lepton Physics}}, 10, 1992
  [\href{https://arxiv.org/abs/hep-ph/9211203}{{\ttfamily hep-ph/9211203}}].

\bibitem{Abel:1992kz}
S.A.~Abel, M.~Dittmar and H.K.~Dreiner, \emph{{Testing locality at colliders
  via Bell's inequality?}},
  \href{https://doi.org/10.1016/0370-2693(92)90071-B}{\emph{Phys. Lett. B}
  {\bfseries 280} (1992) 304}.

\bibitem{Peres:1996dw}
A.~Peres, \emph{{Separability criterion for density matrices}},
  \href{https://doi.org/10.1103/PhysRevLett.77.1413}{\emph{Phys. Rev. Lett.}
  {\bfseries 77} (1996) 1413}
  [\href{https://arxiv.org/abs/quant-ph/9604005}{{\ttfamily
  quant-ph/9604005}}].

\bibitem{Horodecki:1997vt}
P.~Horodecki, \emph{{Separability criterion and inseparable mixed states with
  positive partial transposition}},
  \href{https://doi.org/10.1016/S0375-9601(97)00416-7}{\emph{Phys. Lett. A}
  {\bfseries 232} (1997) 333}
  [\href{https://arxiv.org/abs/quant-ph/9703004}{{\ttfamily
  quant-ph/9703004}}].

\bibitem{Wootters:1997id}
W.K.~Wootters, \emph{{Entanglement of formation of an arbitrary state of two
  qubits}}, \href{https://doi.org/10.1103/PhysRevLett.80.2245}{\emph{Phys. Rev.
  Lett.} {\bfseries 80} (1998) 2245}
  [\href{https://arxiv.org/abs/quant-ph/9709029}{{\ttfamily
  quant-ph/9709029}}].

\bibitem{Clauser:1969ny}
J.F.~Clauser, M.A.~Horne, A.~Shimony and R.A.~Holt, \emph{{Proposed experiment
  to test local hidden variable theories}},
  \href{https://doi.org/10.1103/PhysRevLett.23.880}{\emph{Phys. Rev. Lett.}
  {\bfseries 23} (1969) 880}.

\bibitem{Horodecki1995ViolatingBI}
R.~Horodecki, P.~Horodecki and M.~Horodecki, \emph{Violating bell inequality by
  mixed spin- \{1\}/\{2\} states: necessary and sufficient condition},
  {\emph{Physics Letters A} {\bfseries 200} (1995) 340}.

\bibitem{Bernreuther:2004jv}
W.~Bernreuther, A.~Brandenburg, Z.G.~Si and P.~Uwer, \emph{{Top quark pair
  production and decay at hadron colliders}},
  \href{https://doi.org/10.1016/j.nuclphysb.2004.04.019}{\emph{Nucl. Phys. B}
  {\bfseries 690} (2004) 81}
  [\href{https://arxiv.org/abs/hep-ph/0403035}{{\ttfamily hep-ph/0403035}}].

\bibitem{ATLAS:2020aln}
{\scshape ATLAS} collaboration, \emph{{Measurement of the $t\bar{t}$ production
  cross-section in the lepton+jets channel at $\sqrt{s}=13$ TeV with the ATLAS
  experiment}},
  \href{https://doi.org/10.1016/j.physletb.2020.135797}{\emph{Phys. Lett. B}
  {\bfseries 810} (2020) 135797}
  [\href{https://arxiv.org/abs/2006.13076}{{\ttfamily 2006.13076}}].

\bibitem{CMS:2023qyl}
{\scshape CMS} collaboration, \emph{{First measurement of the top quark pair
  production cross section in proton-proton collisions at $\sqrt{s}$ = 13.6
  TeV}},  \href{https://arxiv.org/abs/2303.10680}{{\ttfamily 2303.10680}}.

\bibitem{ATLAS:2022xfj}
{\scshape ATLAS} collaboration, \emph{{Measurements of differential
  cross-sections in top-quark pair events with a high transverse momentum top
  quark and limits on beyond the Standard Model contributions to top-quark pair
  production with the ATLAS detector at $ \sqrt{s} $ = 13 TeV}},
  \href{https://doi.org/10.1007/JHEP06(2022)063}{\emph{JHEP} {\bfseries 06}
  (2022) 063} [\href{https://arxiv.org/abs/2202.12134}{{\ttfamily
  2202.12134}}].

\bibitem{CMS:2021vhb}
{\scshape CMS} collaboration, \emph{{Measurement of differential $t \bar t$
  production cross sections in the full kinematic range using lepton+jets
  events from proton-proton collisions at $\sqrt {s}$ = 13\,\,TeV}},
  \href{https://doi.org/10.1103/PhysRevD.104.092013}{\emph{Phys. Rev. D}
  {\bfseries 104} (2021) 092013}
  [\href{https://arxiv.org/abs/2108.02803}{{\ttfamily 2108.02803}}].

\bibitem{Uwer:2004vp}
P.~Uwer, \emph{{Maximizing the spin correlation of top quark pairs produced at
  the Large Hadron Collider}},
  \href{https://doi.org/10.1016/j.physletb.2005.01.005}{\emph{Phys. Lett. B}
  {\bfseries 609} (2005) 271}
  [\href{https://arxiv.org/abs/hep-ph/0412097}{{\ttfamily hep-ph/0412097}}].

\bibitem{Bernreuther:2008ju}
W.~Bernreuther, \emph{{Top quark physics at the LHC}},
  \href{https://doi.org/10.1088/0954-3899/35/8/083001}{\emph{J. Phys. G}
  {\bfseries 35} (2008) 083001}
  [\href{https://arxiv.org/abs/0805.1333}{{\ttfamily 0805.1333}}].

\bibitem{Kun}
K.~Cheng, T.~Han and M.~Low, \emph{{Optimizing Entanglement and Bell Inequality
  Violation in Top Anti-Top Events}},  \href{https://arxiv.org/abs/to
  appear}{{\ttfamily to appear}}.

\bibitem{Fraser:2018ieu}
K.~Fraser and M.D.~Schwartz, \emph{{Jet Charge and Machine Learning}},
  \href{https://doi.org/10.1007/JHEP10(2018)093}{\emph{JHEP} {\bfseries 10}
  (2018) 093} [\href{https://arxiv.org/abs/1803.08066}{{\ttfamily
  1803.08066}}].

\bibitem{Kang:2023ptt}
Z.-B.~Kang, A.J.~Larkoski and J.~Yang, \emph{{Towards a Nonperturbative
  Formulation of the Jet Charge}},
  \href{https://doi.org/10.1103/PhysRevLett.130.151901}{\emph{Phys. Rev. Lett.}
  {\bfseries 130} (2023) 151901}
  [\href{https://arxiv.org/abs/2301.09649}{{\ttfamily 2301.09649}}].

\bibitem{Tweedie:2014yda}
B.~Tweedie, \emph{{Better Hadronic Top Quark Polarimetry}},
  \href{https://doi.org/10.1103/PhysRevD.90.094010}{\emph{Phys. Rev. D}
  {\bfseries 90} (2014) 094010}
  [\href{https://arxiv.org/abs/1401.3021}{{\ttfamily 1401.3021}}].

\bibitem{Bernreuther:2015yna}
W.~Bernreuther, D.~Heisler and Z.-G.~Si, \emph{{A set of top quark spin
  correlation and polarization observables for the LHC: Standard Model
  predictions and new physics contributions}},
  \href{https://doi.org/10.1007/JHEP12(2015)026}{\emph{JHEP} {\bfseries 12}
  (2015) 026} [\href{https://arxiv.org/abs/1508.05271}{{\ttfamily
  1508.05271}}].

\bibitem{CMS:2019nrx}
{\scshape CMS} collaboration, \emph{{Measurement of the top quark polarization
  and $\mathrm{t\bar{t}}$ spin correlations using dilepton final states in
  proton-proton collisions at $\sqrt{s} =$ 13 TeV}},
  \href{https://doi.org/10.1103/PhysRevD.100.072002}{\emph{Phys. Rev. D}
  {\bfseries 100} (2019) 072002}
  [\href{https://arxiv.org/abs/1907.03729}{{\ttfamily 1907.03729}}].

\bibitem{ATLAS-CONF-2023-069}
{\scshape ATLAS} collaboration, \emph{{Observation of quantum entanglement in
  top-quark pair production using $pp$ collisions of $\sqrt{s} = 13$~ TeV with
  the ATLAS detector}},  CERN, Geneva (2023).

\bibitem{ParticleDataGroup:2022pth}
{\scshape Particle Data Group} collaboration, \emph{{Review of Particle
  Physics}}, \href{https://doi.org/10.1093/ptep/ptac097}{\emph{PTEP} {\bfseries
  2022} (2022) 083C01}.

\bibitem{Alwall:2011uj}
J.~Alwall, M.~Herquet, F.~Maltoni, O.~Mattelaer and T.~Stelzer, \emph{{MadGraph
  5 : Going Beyond}},
  \href{https://doi.org/10.1007/JHEP06(2011)128}{\emph{JHEP} {\bfseries 06}
  (2011) 128} [\href{https://arxiv.org/abs/1106.0522}{{\ttfamily 1106.0522}}].

\bibitem{Ball:2013hta}
{\scshape NNPDF} collaboration, \emph{{Parton distributions with QED
  corrections}},
  \href{https://doi.org/10.1016/j.nuclphysb.2013.10.010}{\emph{Nucl. Phys. B}
  {\bfseries 877} (2013) 290}
  [\href{https://arxiv.org/abs/1308.0598}{{\ttfamily 1308.0598}}].

\bibitem{Artoisenet:2012st}
P.~Artoisenet, R.~Frederix, O.~Mattelaer and R.~Rietkerk, \emph{{Automatic
  spin-entangled decays of heavy resonances in Monte Carlo simulations}},
  \href{https://doi.org/10.1007/JHEP03(2013)015}{\emph{JHEP} {\bfseries 03}
  (2013) 015} [\href{https://arxiv.org/abs/1212.3460}{{\ttfamily 1212.3460}}].

\bibitem{Czakon:2011xx}
M.~Czakon and A.~Mitov, \emph{{Top++: A Program for the Calculation of the
  Top-Pair Cross-Section at Hadron Colliders}},
  \href{https://doi.org/10.1016/j.cpc.2014.06.021}{\emph{Comput. Phys. Commun.}
  {\bfseries 185} (2014) 2930}
  [\href{https://arxiv.org/abs/1112.5675}{{\ttfamily 1112.5675}}].

\bibitem{Sjostrand:2014zea}
T.~Sj\"ostrand, S.~Ask, J.R.~Christiansen, R.~Corke, N.~Desai, P.~Ilten et~al.,
  \emph{{An introduction to PYTHIA 8.2}},
  \href{https://doi.org/10.1016/j.cpc.2015.01.024}{\emph{Comput. Phys. Commun.}
  {\bfseries 191} (2015) 159}
  [\href{https://arxiv.org/abs/1410.3012}{{\ttfamily 1410.3012}}].

\bibitem{deFavereau:2013fsa}
{\scshape DELPHES 3} collaboration, \emph{{DELPHES 3, A modular framework for
  fast simulation of a generic collider experiment}},
  \href{https://doi.org/10.1007/JHEP02(2014)057}{\emph{JHEP} {\bfseries 02}
  (2014) 057} [\href{https://arxiv.org/abs/1307.6346}{{\ttfamily 1307.6346}}].

\bibitem{Cacciari:2008gp}
M.~Cacciari, G.P.~Salam and G.~Soyez, \emph{{The anti-$k_t$ jet clustering
  algorithm}}, \href{https://doi.org/10.1088/1126-6708/2008/04/063}{\emph{JHEP}
  {\bfseries 04} (2008) 063} [\href{https://arxiv.org/abs/0802.1189}{{\ttfamily
  0802.1189}}].

\bibitem{compare_reconstruct}
J.~Kvita, \emph{Study of methods of resolved top quark reconstruction in
  semileptonic t t- decay: Erratum},
  \href{https://doi.org/10.1016/j.nima.2022.167172}{\emph{Nuclear Instruments
  and Methods in Physics Research Section A: Accelerators, Spectrometers,
  Detectors and Associated Equipment} {\bfseries 1040} (2022) 167172}.

\bibitem{ATLAS_reconstruct}
G.~Aad, B.~Abbott, J.~Abdallah, S.~Khalek, O.~Abdinov, R.~Aben et~al.,
  \emph{Differential top-antitop cross-section measurements as a function of
  observables constructed from final-state particles using pp collisions at s =
  7 tev in the atlas detector},
  \href{https://doi.org/10.1007/JHEP06(2015)100}{\emph{Journal of High Energy
  Physics} {\bfseries 2015} (2015) }.

\bibitem{Collaboration:2267573}
{\scshape CMS} collaboration, \emph{{Object definitions for top quark analyses
  at the particle level}},  CERN, Geneva (2017).

\bibitem{Krohn:2009th}
D.~Krohn, J.~Thaler and L.-T.~Wang, \emph{{Jet Trimming}},
  \href{https://doi.org/10.1007/JHEP02(2010)084}{\emph{JHEP} {\bfseries 02}
  (2010) 084} [\href{https://arxiv.org/abs/0912.1342}{{\ttfamily 0912.1342}}].

\bibitem{dagostini2010improved}
G.~D'Agostini, \emph{Improved iterative bayesian unfolding},  2010.

\bibitem{Hocker:1995kb}
A.~Hocker and V.~Kartvelishvili, \emph{{SVD approach to data unfolding}},
  \href{https://doi.org/10.1016/0168-9002(95)01478-0}{\emph{Nucl. Instrum.
  Meth. A} {\bfseries 372} (1996) 469}
  [\href{https://arxiv.org/abs/hep-ph/9509307}{{\ttfamily hep-ph/9509307}}].

\bibitem{Adye:2011gm}
T.~Adye, \emph{{Unfolding algorithms and tests using RooUnfold}},  in
  \emph{{PHYSTAT 2011}}, (Geneva), pp.~313--318, CERN, 2011,
  \href{https://doi.org/10.5170/CERN-2011-006.313}{DOI}
  [\href{https://arxiv.org/abs/1105.1160}{{\ttfamily 1105.1160}}].

\bibitem{Stanley_2022}
M.~Stanley, P.~Patil and M.~Kuusela, \emph{Uncertainty quantification for
  wide-bin unfolding: one-at-a-time strict bounds and prior-optimized
  confidence intervals},
  \href{https://doi.org/10.1088/1748-0221/17/10/P10013}{\emph{Journal of
  Instrumentation} {\bfseries 17} (2022) P10013}.

\bibitem{Blobel:1984ku}
V.~Blobel, \emph{{Unfolding Methods in High-energy Physics Experiments}},  in
  \emph{{1984 CERN School of Computing}}, 12, 1984.

\bibitem{Blobel:2011fih}
V.~Blobel, \emph{{Unfolding Methods in Particle Physics}},  in \emph{{PHYSTAT
  2011}}, (Geneva), pp.~240--251, CERN, 2011,
  \href{https://doi.org/10.5170/CERN-2011-006.240}{DOI}.

\bibitem{Schmitt:2016orm}
S.~Schmitt, \emph{{Data Unfolding Methods in High Energy Physics}},
  \href{https://doi.org/10.1051/epjconf/201713711008}{\emph{EPJ Web Conf.}
  {\bfseries 137} (2017) 11008}
  [\href{https://arxiv.org/abs/1611.01927}{{\ttfamily 1611.01927}}].

\bibitem{kuusela2012statistical}
M.~Kuusela et~al., \emph{Statistical issues in unfolding methods for high
  energy physics},  Master's thesis, Aalto University, 2012.

\bibitem{Brandenburg:2002xr}
A.~Brandenburg, Z.G.~Si and P.~Uwer, \emph{{QCD corrected spin analyzing power
  of jets in decays of polarized top quarks}},
  \href{https://doi.org/10.1016/S0370-2693(02)02098-1}{\emph{Phys. Lett. B}
  {\bfseries 539} (2002) 235}
  [\href{https://arxiv.org/abs/hep-ph/0205023}{{\ttfamily hep-ph/0205023}}].

\bibitem{ATLAS:2018mgv}
{\scshape ATLAS} collaboration, \emph{{Search for the Decay of the Higgs Boson
  to Charm Quarks with the ATLAS Experiment}},
  \href{https://doi.org/10.1103/PhysRevLett.120.211802}{\emph{Phys. Rev. Lett.}
  {\bfseries 120} (2018) 211802}
  [\href{https://arxiv.org/abs/1802.04329}{{\ttfamily 1802.04329}}].

\bibitem{CMS:2019hve}
{\scshape CMS} collaboration, \emph{{A search for the standard model Higgs
  boson decaying to charm quarks}},
  \href{https://doi.org/10.1007/JHEP03(2020)131}{\emph{JHEP} {\bfseries 03}
  (2020) 131} [\href{https://arxiv.org/abs/1912.01662}{{\ttfamily
  1912.01662}}].

\end{thebibliography}\endgroup
\end{document}